\documentclass{emulateapj}
\usepackage[bottom]{footmisc}
\usepackage{color,amsmath,enumitem}%,breqn}

\shorttitle{Bayesian Analysis of Multiple Populations}
\shortauthors{Wagner-Kaiser et al.}

\begin{document}

\newcommand{\markup}[1]{\textbf{\textcolor{red}{{\fontfamily{garamond}\selectfont{#1}}}}}
\renewcommand*{\thefootnote}{\fnsymbol{footnote}}

% New Commands
\newcommand{\bmu}{\mbox{\boldmath{$\mu$}}}
\newcommand{\balpha}{\mbox{\boldmath{$\alpha$}}}
\newcommand{\bSigma}{\mbox{\boldmath{$\Sigma$} }}
\newcommand{\bTheta}{\mbox{\boldmath{$\Theta$}}}
\newcommand{\bOmega}{\mbox{\boldmath{$\Omega$}}}
\newcommand{\btheta}{\mbox{\boldmath{$\theta$}}}
\newcommand{\bphi}{\mbox{\boldmath{$\phi$}}}
\newcommand{\bPhi}{\mbox{\boldmath{$\Phi$}}}
\newcommand{\bM}{\mbox{\boldmath{$M$}}}
\newcommand{\bR}{\mbox{\boldmath{$R$}}}
\newcommand{\bZ}{\mbox{\boldmath{$Z$ }}}
\newcommand{\bX}{\mbox{\boldmath{$X$ }}}
\newcommand{\bG}{\mbox{\boldmath{$G$}}}
\newcommand{\bF}{\mbox{\boldmath{$F$}}}
\newcommand{\bV}{\mbox{\boldmath{$V$}}}
\newcommand{\bGMS}{\mbox{\boldmath{$G$}}_{\textrm{MS/RG}}}
\newcommand{\bGWD}{\mbox{\boldmath{$G$}}_{\textrm{WD}}}
\newcommand{\bY}{\mbox{\boldmath{$\Psi$}}}
\newcommand{\diag}{\textrm{diag}}
\newcommand{\thetaage}{\mbox{$\theta_{\text{age}}$}}
\newcommand{\thetaFeH}{\mbox{$\theta_{\text{[Fe/H]}}$}}
\newcommand{\thetamod}{\mbox{$\theta_{m - M_V}$}}
\newcommand{\thetaabs}{\mbox{$\theta_{A_V}$}}
\newcommand{\thetaY}{\mbox{$\theta_{Y}$}}
\newcommand{\thetaC}{\mbox{$\theta_{C}$}}
\newcommand{\phiY}{\mbox{$\phi_{Y}$}}
\newcommand{\phiprog}{\mbox{$\phi_{\text{prog\,age}}$}}
\newcommand{\phiTeff}{\mbox{$\phi_{T_{\text{eff}}}$}}
\newcommand{\phiradius}{\mbox{$\phi_\text{radius}$}}
\newcommand{\philogg}{\mbox{$\phi_{\log g}$}}
\newcommand{\bxi}{\mbox{\boldmath{$\Xi$}}}
\newcommand{\bDelta}{\mbox{\boldmath{$\Delta$}}}
\newcommand{\indep}{\stackrel{indep}{\sim}}
\newcommand{\lambdak}{\mbox{$\lambda_{k}$}}

\title{Bayesian Analysis of Two Stellar Populations in Galactic Globular Clusters II: NGC 5024, NGC 5272, and NGC 6352}
\author{R. Wagner-Kaiser*$^{1}$, D. C. Stenning$^{2}$, E. Robinson$^{3}$, T. von Hippel$^{4}$, A. Sarajedini$^{1}$, D. A. van Dyk$^{5}$, N. Stein$^{6}$, W. H. Jefferys$^{7, 8}$}
\altaffiltext{1}{Bryant Space Center, University of Florida, Gainesville, FL}
\altaffiltext{2}{Sorbonne Universit\'{e}s, UPMC-CNRS, UMR 7095, Institut d'Astrophysique de Paris, F-75014 Paris, France}
\altaffiltext{3}{Argiope Technical Solutions, FL}
\altaffiltext{4}{Center for Space and Atmospheric Research, Embry-Riddle Aeronautical University, Daytona Beach, FL}
\altaffiltext{5}{Imperial College London, London, UK}
\altaffiltext{6}{The Wharton School, University of Pennsylvania, Philadelphia, PA}
\altaffiltext{7}{University of Texas, Austin, TX, USA}
\altaffiltext{8}{University of Vermont, Burlington, VT, USA}
\email{*rawagnerkaiser@astro.ufl.edu}

%%%%%%%%%%%%%%%%%%%%%%%%%%%%%%%%%%%%%%%%%%%%%%%%%%%%%%%%%%%%%%%%%%%%%%%%%%%%%% 
%                                                                                                                                                ABSTRACT
%%%%%%%%%%%%%%%%%%%%%%%%%%%%%%%%%%%%%%%%%%%%%%%%%%%%%%%%%%%%%%%%%%%%%%%%%%%%%%

\begin{abstract}
We use Cycle 21 Hubble Space Telescope (HST) observations and HST archival ACS Treasury observations of Galactic Globular Clusters to find and characterize two stellar populations in NGC 5024 (M53), NGC 5272 (M3), and NGC 6352. For these three clusters, both single and double-population analyses are used to determine a best fit isochrone(s). We employ a sophisticated Bayesian analysis technique to simultaneously fit the cluster parameters (age, distance, absorption, and metallicity) that characterize each cluster. For the two-population analysis, unique population level helium values are also fit to each distinct population of the cluster and the relative proportions of the populations are determined. We find differences in helium ranging from $\sim$0.05 to 0.11 for these three clusters. Model grids with solar $\alpha$-element abundances ([$\alpha$/Fe] =0.0) and enhanced $\alpha$-elements ([$\alpha$/Fe]=0.4) are adopted. 
\end{abstract}

%%%%%%%%%%%%%%%%%%%%%%%%%%%%%%%%%%%%%%%%%%%%%%%%%%%%%%%%%%%%%%%%%%%%%%%%%%%%%% 
%                                                                                                                                                INTRODUCTION
%%%%%%%%%%%%%%%%%%%%%%%%%%%%%%%%%%%%%%%%%%%%%%%%%%%%%%%%%%%%%%%%%%%%%%%%%%%%%%

\section{Introduction}\label{Intro}

Our understanding of globular clusters, their role in the formation and merger history of the Milky Way, as well as their role in other galaxies rests largely on their analyses as single stellar populations. The classical view of globular clusters describes a group of stars all born of the same material at the same time. This perspective implies that all stars should overall be co-eval, co-spatial, and iso-metallic. Under the assumption of a simple, single population, globular clusters have long been used as fossils to trace the kinematic and chemical evolution of galaxies.

However, in the past decade or so, the assumptions about globular clusters have been called into question. Although the evidence has been amassing for decades, recent studies have found overwhelming evidence that globular clusters harbor more than one distinct population of stars (\citealt{Bedin:2004}, \citealt{Gratton:2004}, \citealt{Carretta:2006}, \citealt{Villanova:2007}, \citealt{Piotto:2007}, \citealt{Piotto:2009}, \citealt{Milone:2009}, \citealt{Milone:2012a}, among others). \cite{Cannon:1973} were first to find traces of multiple populations in globular clusters when taking photoelectric photometry of the massive cluster $\omega$ Centauri. They found that the red giant branch had a spread larger than could be explained by the apparent effects of photometric error, field stars, and reddening. Further spectral studies found similar results in a number of globular clusters, but it wasn't until \cite{Lee:1999} that unequivocal photometric evidence for multiple populations was obtained of $\omega$ Cen using the Hubble Space Telescope. Shortly thereafter, \cite{Bedin:2004} observed $\omega$ Cen with deep, high-precision HST images, exposing a variety of intricacies never seen before in a globular cluster. More recent studies (\citealt{Piotto:2007}, \citealt{Milone:2009}, \citealt{Milone:2012}, \citealt{Milone:2012a}, \citealt{Milone:2013}, \citealt{Nardiello:2015}, \citealt{Piotto:2015}) demonstrate that many globular clusters have undergone more than one epoch of star formation, a conclusion in dramatic contrast with the classical ``simple stellar population" hypothesis.

It now appears that most, if not all, globular clusters contain multiple populations of stars (\citealt{Piotto:2015}). Of the well-studied clusters thus far, all multiple population attributes appear to manifest in unique ways (different combinations of varying helium abundances, number of populations, proportions of stars belonging to each population, etc.). Nonetheless, most studies agree that in many clusters, helium likely drives differences of multiple populations observed in color-magnitude diagrams (CMDs), primarily on the main sequence and red giant branch (\citealt{Gratton:2012}). One popular picture of formation includes a second generation of stars, and sometimes subsequent generations, enriched by processed material from intermediate-mass stars (\citealt{Renzini:2008}). The enriched ejecta from the first generation of intermediate mass stars then gathers in the central regions of the cluster, due to the gravitational potential well, where the second generation of stars can then form. The second generation of stars has enriched helium content with respect to the first generation, as well as differences in the light abundances. Hence, further generations of stars (third, fourth, etc.) follow a similar scenario and have even more enhanced abundances. Other possible scenarios suggest accretion onto proto-planetary disks or extremely massive stars; however, there is currently no proffered scenario that is able to explain the wide range of abundance patterns that are currently observed (\citealt{Bastian:2015}).

The differences in abundances lead to observable differences at ultraviolet wavelengths, which have an incredible potential to help us study multiple populations. Ultraviolet filters are sensitive to variations in particular metals, specifically carbon, oxygen, and nitrogen - the elements that we expect to indicate chemical enrichment and gas recycling. Although photometry cannot provide accuracy of elemental abundances that rivals spectroscopy, photometry provides a huge gain in the sheer numbers of stars that can be attributed to distinct populations.

Here, we characterize two stellar populations of the globular clusters NGC 5024, NGC 5272, and NGC 6352 by adopting a Bayesian approach for model fitting. For a long time, isochrones have been fit by hand to data, by choosing a model and adjusting parameters until a fit ``looks good". In the past, numerical approaches, including Bayesian techniques, have occasionally been used to find a best fit (\citealt{Andreuzzi:2011}, \citealt{Jorgensen:2005}, \citealt{Naylor:2006}, \citealt{Hernandez:2008}, \citealt{Janes:2013}, cite alt{de-Souza:2015}). With our Bayesian approach, we can interpolate to high precision using a grid of isochrones and more reliably determine the most likely isochrone fit for a cluster, via the posterior distribution of the unknown cluster parameters (\citealt{Jeffery:2016}). This enables us to provide principled measures of uncertainty on both the fitted parameters and the fitted isochrones. The Bayesian Analysis for Stellar Evolution software suite (BASE-9; see \citealt{von-Hippel:2006}, \citealt{De-Gennaro:2009}, \citealt{van-Dyk:2009}, \citealt{Stein:2013}), fits a cluster consisting of a single population by deriving the joint posterior distribution of age, distance, absorption, metallicity, and the initial stellar masses, while allowing for field star contamination. We have adapted BASE-9 for use on a globular cluster assumed to host two distinct populations (\citealt{Stenning:2016}). With theoretical models, we are able to precisely characterize two helium values in clusters and determine the proportion of stars in each population. We use Cycle 21 data from HST (GO Cycle 21 Proposal 13297; \cite{Piotto:2015}) of Galactic globular clusters in the UVIS filters (F275W, F336W, F438W) in conjunction with archival ACS Treasury data (F606W and F814W filters; GO Cycle 14 Proposal 10775; \citealt{Sarajedini:2007}) to achieve this goal.

Almost ubiquitously, Galactic globular clusters have now been observed to consist of more than one population, but they exhibit an inexplicable range of patterns. By beginning to analyze helium abundances in a growing number of clusters using BASE-9 (as described in \citealt{Stenning:2016}), we hope to learn more about multiple populations in globular clusters. In doing so, we gain insight into the formation mechanism of these clusters and the implications for the history of the Milky Way. In Section \ref{Data}, we discuss the HST data we use to achieve our goals and in Section \ref{Methods}, we detail the Bayesian analysis technique. In Section \ref{SinglePop}, we present results of a single population analysis of our clusters. In Section \ref{MultiPop}, the results of analyzing the same clusters with ultraviolet photometry and a two-population analysis are presented. We compare the BASE-9 single population results to double population results in Section \ref{Discussion}, and interpret these in context of previous studies of these clusters. We conclude in Section \ref{Conclusion}.

%%%%%%%%%%%%%%%%%%%%%%%%%%%%%%%%%%%%%%%%%%%%%%%%%%%%%%%%%%%%%%%%%%%%%%%%%%%%%% 
%                                                                                                                                                DATA
%%%%%%%%%%%%%%%%%%%%%%%%%%%%%%%%%%%%%%%%%%%%%%%%%%%%%%%%%%%%%%%%%%%%%%%%%%%%%%

\section{Data}\label{Data}

The clusters in our sample were previously observed by the ACS Globular Cluster Treasury Survey (GO Cycle 14 Proposal 10775; PI: Sarajedini). The ACS Globular Cluster Treasury observed 65 clusters in the HST F606W and F814W filters (\citealt{Sarajedini:2007}, \citealt{Anderson:2008}; 71 clusters including \citealt{Dotter:2011}). These observations provided a wealth of data and new insights into the Galactic globular cluster population.

The new data from HST Cycle 21 Proposal GO 13297 (PI: Piotto) obtained 131 orbits for 47 of the globular clusters in the ACS Treasury survey. The observations extend the wavelength coverage into the ultraviolet with the F275W, F336W, and F438W filters. These passbands disentangle multiple populations in globular clusters due to their sensitivity to C, N, and O abundances. Specifically, the F275W filter contains an OH band, the F336W filter contains an NH band, and the F438W filter contains both CN and CH bands. These filters distinguish among different CNO contents correlated with helium, and thus are able to separate populations in color-magnitude space.

The intermediate level photometry (see \citealt{Piotto:2015} for details) provide a unified star list for the F275W, F336W, and F438W filters as well as the F606W and F814W filters from the ACS Globular Cluster Treasury Survey (\citealt{Sarajedini:2007}). We provide a summary of the three clusters analyzed in this work in Table \ref{clusterlist}. 

\begin{figure*}
\plotone{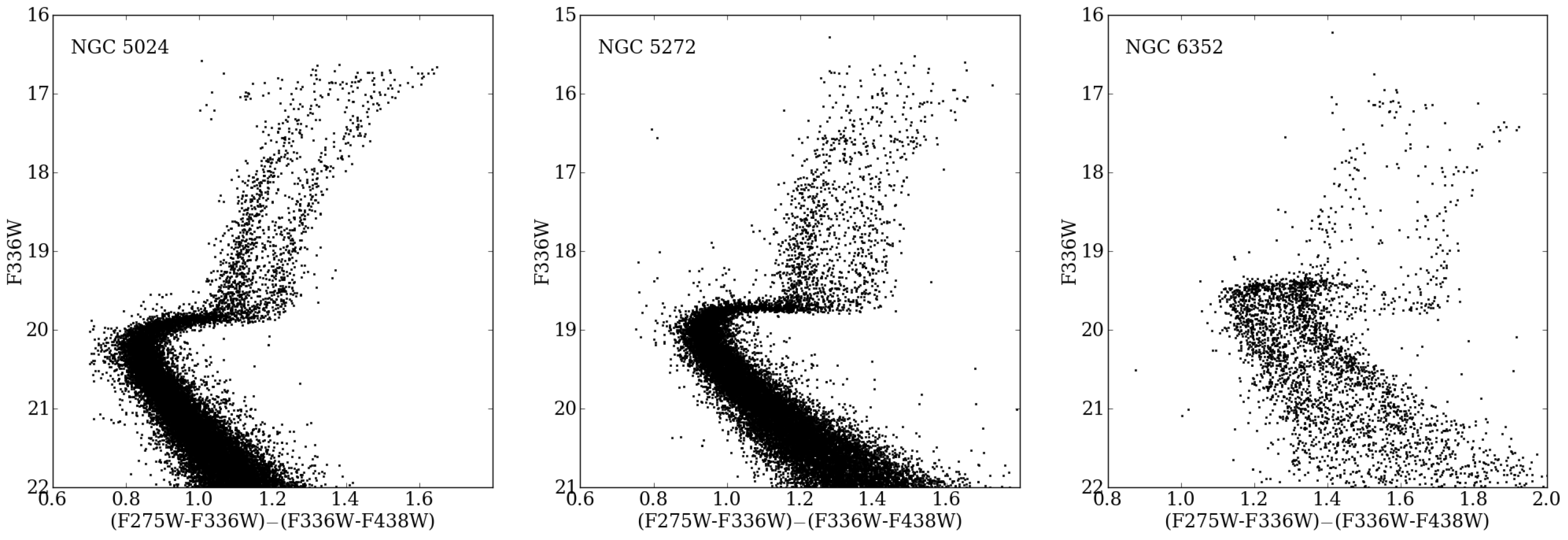}
\caption{Ultraviolet CMDs of the clusters in our study. The color on the x-axis is chosen as a combination of the three ultraviolet filters to maximize the separation of the populations in the CMD.} % Left: NGC 5024; middle: NGC 5272; right: NGC 6352.
\label{3CMDs}
\end{figure*}

We use the seven-year baseline between the HST ACS Cycle 14 photometry and the HST UVIS Cycle 21 photometry to remove many non-cluster stars based on their pixel location errors (remaining field stars are taken into account in our model; \citealt{Stenning:2016}). Additionally, we use photometry quality flags to reject poor photometry (\citealt{Piotto:2015}). We also remove horizontal branch (HB) stars from the samples because they are not included in the theoretical models that we use. CMDs of the three clusters NGC 5024, NGC 5272, and NGC 6352 are shown in Figure \ref{3CMDs}, where the x-axis is chosen as a combination of the HST ultraviolet filters to visually accentuate the bi-modality of populations in these clusters.

\begin{table*}
\centering
%    \begin{minipage}{180mm}
    \caption{Summary of Cluster Sample$^a$}
    \begin{tabular}{@{}cccccc@{}}
    \hline
 \textbf{Name} &  \textbf{Right Ascension} & \textbf{Declination}  & \textbf{[Fe/H]}  & \textbf{Distance Modulus}  & \textbf{E(B-V)}\\  
\hline
NGC5024 & 13$^{h}$ 12$^{m}$ 55$^{s}$ & +18$^{\circ}$ 10' 05" & --2.1 & 16.32 & 0.02 \\
NGC5272 & 13$^{h}$ 42$^{m}$ 12$^{s}$ & +28$^{\circ}$ 22' 38" & --1.5 & 15.07 & 0.01 \\
NGC6352 & 17$^{h}$ 25$^{m}$ 29$^{s}$ & --48$^{\circ}$ 25' 20" & --0.70\footnotemark[2] & 14.43 & 0.22 \\ \hline
    \end{tabular}
%    \end{minipage}
   \label{clusterlist}
   \footnotetext[1]{Data from \cite{Harris:2010} unless otherwise noted.}
   \footnotetext[2]{From \cite{Roediger:2014}.}
\end{table*}

\subsection{Photometric Errors}\label{ErrorSection}
\begin{figure} 
\epsscale{1.25}
\plotone{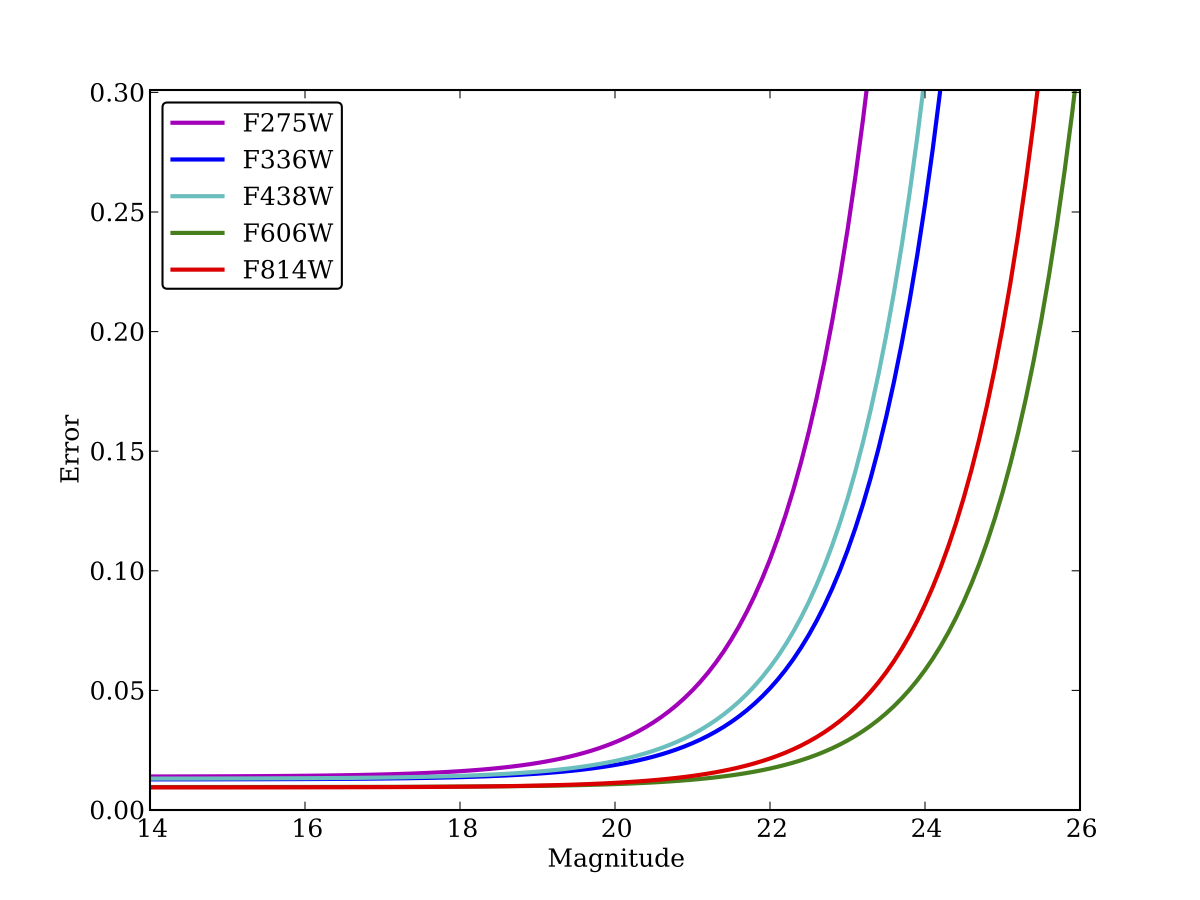}
\caption{The error profiles used to generate photometric errors for each observed star in each filter. A minimum photometric error of 0.01 is used as a conservative lower limit.}
\label{ErrorProfile}
\end{figure}

Photometric errors are not currently available for these UV data. Eventually, artificial star tests will provide a valuable estimate of the photometric uncertainty for each observed star. Any principled statistical analysis requires measurement errors for all photometry. As such, some rough, reasonable estimate of the error of each datum is necessary. We use the HST exposure time calculator to estimate errors based on magnitude and filter, and adopt a reasonable minimum error of 0.01 mag. The error profiles we use are shown in Figure \ref{ErrorProfile}.

Without artificial star tests, we cannot know in particular which stars have higher photometric uncertainties (for instance, due to proximity to a brighter star). However, the lower limit of a 0.01 mag error leads to practical errors that increase with magnitude, allowing the Bayesian framework to take into account the knowledge that fainter stars are more difficult to accurately measure. For example, in the case of NGC 5024, the error distribution peaks at 0.035 in the F275W filter, 0.025 in the F336W filter, 0.03 in F438W, and 0.015 and 0.02 respectively in F606W and F814W. This reflects the actual errors we expect to eventually see post-artificial star tests.

\subsection{NGC 5024 (M53)}\label{NGC5024back}

The metal-poor globular cluster NGC 5024 is thought to be a possible former member of Sagittarius (\citealt{Chun:2010}), making it an interesting target in which to examine multiple populations. If the characteristics of multiple populations in NGC 5024 are significantly different from those of the Galactic globular clusters, it may imply a distinct formation or enrichment scenario in different environments.

NGC 5024 has not had extensive studies of its multiple populations, but \cite{Caloi:2011} suggest that NGC 5024 is a primarily first population cluster, due to the short HB in the CMD and its classification as a Oosterhoff type-II cluster. They require a heavily enriched helium value 0.42 to reproduce the blue hook of the HB, using the same models as \cite{di-Criscienzo:2011}. However, \cite{Jang:2014} suggests this cluster should be primarily second population stars based on their modeling of the Oosterhoff dichotomy. With our analysis, we weigh in on the helium values of both populations as well as the relative proportions.

\subsection{NGC 5272 (M3)}\label{NGC5272back}

As one of the largest, brightest, most highly-populated clusters, NGC 5272 has been the target of many studies. It is known to have hundreds of variable stars, providing a rich population of horizontal branch stars to study.

\cite{Jurcsik:2003} found a large spread in the spectroscopic metallicities of NGC 5272 horizontal branch stars, suggesting an internal spread of iron abundances, which has been found in other massive clusters such as $\omega$ Cen (\citealt{Johnson:2008}). \cite{Jurcsik:2003} also suggests that the HB stars are 0.3 to 0.4 dex more metal-rich than the RGB stars, which would be unusual. Using high-precision spectra to determine empirical gravities, \cite{Catelan:2009} determined a helium enhancement of only 0.01 is necessary to explain the observed differences between the red and blue HB stars.

Previous studies of the helium in NGC 5272, primarily focusing on the horizontal branch stars, suggest that little ($\Delta$Y$\sim$0.01-0.02) to no helium enhancement is needed to explain the HB morphology (\citealt{Catelan:2009}, \citealt{Valcarce:2010}, \citealt{Dalessandro:2013}). However, as these studies focus on the horizontal branch only, it is unclear what additional information on helium content might be encoded into other parts of the CMD.

\subsection{NGC 6352}\label{NGC6352back}

NGC 6352 is a metal-rich cluster, and typically attributed as a member of the bulge or disk population of GCs based on its kinematics (\citealt{Feltzing:2009}). Metal-rich clusters ([Fe/H]$\textgreater$--1.0) are often difficult to model, and hence fit, with standard isochrones. The cluster also suffers higher extinction than either NGC 5024 or NGC 5272.

Previous studies have shown a clear bi-modality in the strength of CN vs CH bands in main sequence stars (\citealt{Pancino:2010}), suggesting multiple populations. The new HST UVIS photometry confirms their existence.

While \cite{Feltzing:2009} finds an iron abundance of [Fe/H]=--0.55 from high-resolution spectroscopy of 9 HB stars, \cite{Roediger:2014} finds a cluster-averaged [Fe/H] of --0.70 from spectroscopy, which we use as a prior for [Fe/H] (see Section \ref{Methods}). The cluster is thought to be enhanced in [$\alpha$/Fe] to about 0.2 (\citealt{Feltzing:2009}, \citealt{Dotter:2011}).

A recent study by \cite{Nardiello:2015}, also using photometry from the UVIS treasury, finds that the two populations in NGC 6352 differ by $\sim$0.03 in helium abundance through examination of RGB fiducials.

%%%%%%%%%%%%%%%%%%%%%%%%%%%%%%%%%%%%%%%%%%%%%%%%%%%%%%%%%%%%%%%%%%%%%%%%%%%%%% 
%                                                                                                                                                METHODS
%%%%%%%%%%%%%%%%%%%%%%%%%%%%%%%%%%%%%%%%%%%%%%%%%%%%%%%%%%%%%%%%%%%%%%%%%%%%%%

\section{Methods}\label{Methods}

We employ a sophisticated Bayesian statistical method to fit isochrones to these globular clusters for multiple reasons. First, as previously mentioned, our robust statistical approach is both objective and reproducible. Secondly, we have the ability to fit multiple parameters simultaneously and explore their correlations while incorporating individual errors on each data point. Additionally, we use a large isochrone grid and interpolate among the models, which allows us to achieve a greater precision than traditional techniques (\citealt{Jeffery:2016}). The Bayesian approach allows inference of complex non-linear correlations among the sampled parameters, where simple point estimates and standard errors are insufficient (e.g.: \citealt{OMalley:2013}, \citealt{Andreon:2015}). Finally, compared to more standard methods, Bayesian analyses readily provide a posterior probability distribution for each parameter, as opposed to a singular value with standard errors. These posterior distributions can be especially valuable when they are skewed or otherwise non-Gaussian. All of these advantages are particularly important with the added complexity of multiple populations.

Our software suite, BASE-9, was developed to determine cluster parameters for a single population of stars using sophisticated Bayesian techniques, given a set of photometry and theoretical models (\citealt{von-Hippel:2006}, \citealt{De-Gennaro:2009}, \citealt{van-Dyk:2009}, \citealt{Stein:2013}). The software, including the updates developed for this two population study, is available as open source code from GitHub (https://github.com/argiopetech/base/releases) and via executables through Amazon Web Services. Installation and instruction for BASE-9 may be found in the User Manual (\citealt{von-Hippel:2014}).

The single population version of BASE-9 can simultaneously fit a variety of cluster-level parameters (distance, absorption, age, and metallicity) and star-specific parameters (mass, binarity, and cluster membership). In order to examine and characterize populations in globular clusters, we have extended BASE-9 to simultaneously sample population-level parameters, specifically two helium fractions, one for each population. We also estimate the proportion of stars in each population, quantified as the probability that a star chosen from the cluster at random belongs to the population with the lower helium content (\citealt{Stenning:2016}). The details of the statistical model for the two-population Bayesian analysis are laid out in \cite{Stenning:2016}, and we briefly summarize them here. Our expanded statistical model incorporates a hierarchical structuring of the parameters: {\it cluster parameters} are common to all cluster stars (age, distance, metallicity, and absorption), {\it population parameters} are common to stars belonging to a particular population but may differ between populations (helium and the proportion of stars in each population), and {\it stellar parameters} are allowed to vary on a star-to-star basis (mass, binarity, and cluster membership); see Table \ref{tbl:multipop_params}.

We observe photometric magnitudes in $n$ different filters for each of $N$ stars. The photometric magnitudes for star $i$ are contained in the vector $X_i$, with known (independent) photometric errors in the diagonal of matrix $\Sigma_i$. We define indicator variables $\bZ = (Z_1, \ldots, Z_N)$ such that $Z_i = 1$ if star $i$ is a cluster star and $Z_i = 0$ if it is a field star. These indicator variables allow us to separately model cluster stars versus field stars. The observed magnitudes of cluster stars are modeled as $n$-dimensional multivariate Gaussian distributions, such that
\begin{align}
&P(\bX_i | \bSigma_i, M_i,R_i, \bTheta, \phi_{Y}, Z_i=1) =  \nonumber \\
&{1\over \sqrt{(2\pi)^{n}|\bSigma_i|}} \exp\bigg(-{1\over2}\Big(\bX_i - \bmu_i\Big)^\top \bSigma_i^{-1}\Big(\bX_i - \bmu_i\Big)   \bigg). &&
\label{eq:cluster_lik}
\end{align}

where $M_i$ and $R_i$ are the mass and mass ratio of star $i$, respectively, $\Theta$ = ($\thetaage$, $\thetamod$, $\thetaabs$, $\thetaFeH$) are the cluster parameters, and $\thetaY$ is helium. $\bX$ represents the photometric data as a collection of vectors and $\bSigma$ is a collection of matrices defining the photometric errors.

We assume that field stars are uniformly distributed over the range of the data. That is, following \cite{van-Dyk:2009}, we model field star magnitudes as
\begin{equation*}
P(\bX_i | Z_i=0) = c \quad \hbox{ if } \text{min}_j  \leq x_{ij} \leq \text{max}_j, j=1,\ldots, n,
\end{equation*}
over the range of filters $j$ with $c=\left[\prod_{j=1}^n (\max_j -\min_j)\right]^{-1}$ and $P(\bX_i | Z_i=0)$ is everywhere else zero. The field star model does not depend on any of the other parameters (e.g.: $\thetaage$, $\thetaFeH$ , $\thetamod$, etc.).

Following this, we augment the single population likelihood function to allow for two populations each with a unique helium content. \bM, \bR, \bTheta, \bPhi, and \bZ  are vectors of each star $i$ for the masses, mass ratios, cluster parameters (\thetaage, \thetamod, \thetaabs, \thetaFeH), population parameters (helium values $\phi_{YA}$, $\phi_{YB}$ and proportions $\phi_{pA}$, $\phi_{pB}$ for population k=A and k=B), and cluster versus field star membership. The parameters of the hierarchical model are also given in Table \ref{tbl:multipop_params} for reference. The two-population likelihood function is then:

\begin{eqnarray}
&&L(\bM,\bR, \bTheta, \bPhi, \bZ | \bX, \bSigma) \nonumber \\
&& \; = \prod_{i=1}^{N} \bigg[ Z_{i} \times \sum_{kin (A,B)}\phi_{pk}P(\bX_i | \bSigma_i, M_i,R_i, \bTheta, \phi_{Yk}, Z_i=1) \nonumber \\
&& \qquad \qquad \qquad + (1- Z_i) \times P(\bX_i | Z_i=0) \bigg].  \\
\label{eq:multipop_like} \nonumber
\end{eqnarray}

\begin{table*}[t!]
\begin{center}
\caption{Two-Population Model Parameters (From \citealt{Stenning:2016})}
\label{tbl:multipop_params}
\begin{tabular}{llr}
\hline
 {\small Parameter} & {\small Description} & {\small Notation} \\  \hline
	{\bf Cluster Parameters} \\
	{\it Age} & $\log_{10}$ of cluster age in years & \thetaage  \\ 
	{\it Distance} & distance modulus in mag  & \thetamod  \\
	{\it Absorption} & absorption in the V-band in mag & \thetaabs  \\ 
	{\it Metallicity}  & $\log_{10}$ of iron-to-hydrogen ratio relative to Sun in dex \qquad \qquad & \thetaFeH \vspace{0.1cm} \\ \hline
	{\bf Population Parameters} \qquad \qquad \\
	{\it Proportion} &  proportion of stars from a population & $\phi_{pk}$ \\
	{\it Helium Abundance} & mass fraction of helium & $\phi_{Yk}$ \vspace{0.1cm} \\ \hline
	{\bf Stellar Parameters} \\
 	{\it Initial Mass} &  Zero Age Main Sequence mass in solar units, $M_{\odot}$ & $M_i$\\
   %$M_{\odot}$\\
	{\it Mass Ratio} & ratio of secondary to primary initial masses & $R_i$ \\
	{\it Cluster Membership} & indicator for cluster membership & $Z_i$ \vspace{0.1cm} \\
\hline
\end{tabular}
\end{center}
\end{table*}

We use Dartmouth Stellar Evolution Database (DSED) models that span an age range (9 to 15 Gyr), metallicity range (--2.5 to 0.5 dex), helium fraction range (0.23 to 0.40), and different [$\alpha$/Fe] values (0.0 and +0.4). A Markov-chain Monte Carlo (MCMC) algorithm, specifically an adaptive Metropolis (AM) implementation, explores the posterior distribution; see \cite{Stenning:2016} for details. We run each MCMC chain for 11,000 iterations. After the first 1000 iterations, which we discard as burn-in, the chain automatically adapts to the observed correlations in earlier iterations to make sampling more efficient. For each fitted model, we run three MCMC chains using over dispersed starting values and visually inspect a plot of the chains to assess convergence. We also compute the Gelman-Rubin statistic (\citealt{Gelman:1992}) on the 10,000 post burn-in iterations per chain of each parameter in each cluster fit, and all values were below 1.13. \footnote{We used the {\tt gelman.diag} function in the coda R package (with {\tt autoburnin=FALSE}) to compute the Gelman-Rubin statistic, also known as the `potential scale reduction factor'.}

The inputs for BASE-9 include the photometry of the stellar cluster in multiple filters, prior distributions for metallicity, distance, absorption, and helium content. Prior distributions on most of these parameters are assumed to be Gaussian, usually in the log of the quantity, consistent with traditional use (e.g. [Fe/H], (m-M)$_{V}$). or A$_{V}$, we specify a Gaussian distribution truncated at zero. We assume a \cite{Cardelli:1989} R$_{V}$ = 3.1 reddening law. We use an uninformative prior on age that is uniform in log(age) from 1 to 15 Gyr. As we do not have reliable estimates for helium abudances, we use uninformative uniform priors on these quantities, with the prior on Y$_{A}$ uniform from 0.15 to 0.30 and the prior on Y$_{B}$ uniform from 0.15 to 0.45, requiring that Y$_{B}$ $\textgreater$ Y$_{A}$. We also do not have any a priori information regarding the fraction of stars that may belong to each population, and so we use a uniform prior distribution over the range 0 to 1. For metallicity, distance, and absorption we specify Gaussian prior distributions, with means set according to published values and standard deviations chosen to be reasonably conservative (see Tables \ref{NGC5024results} through \ref{NGC6352results}). In particular, we use standard deviations of 0.05 for metallicity (0.025 for spectroscopic values), and 0.05 for distance. For the absorption, although errors are typically 10\% of the absorption value (\citealt{Harris:2010}), we conservatively use one third of the published value as the standard deviation for the truncated Gaussian. The output of BASE-9 is a correlated sample from the joint posterior distribution for distance, age, metallicity, absorption, along with the two population helium fractions, the relative population proportions, and the individual stellar masses. 

Currently running BASE-9 with the binary option with several thousand stars is computationally prohibitive. However, as the three clusters of interest have low binary fractions (5\%, \citealt{Milone:2012b}), treating the stars as single systems should not have a significant effect on the final results.

Although we know these clusters exhibit multiple population qualities, we analyze them both as single and as double populations for comparison. Additionally, we perform our isochrone fits for an [$\alpha$/Fe] enrichment of 0.0 (solar) and +0.4 (enhanced). From the cleaned photometry, as discussed in Section \ref{Data}, we randomly select a subsample of stars, limited to seven magnitudes fainter than the \cite{Harris:2010} distance modulus of each cluster, in order to have a consistent cutoff for every cluster. Several thousand stars is more than sufficient to obtain a robust fit from BASE-9 and using $\lesssim$ 3000 stars decreases the computational time required for each cluster. We randomly select 1500 stars above the main sequence turn-off point (MSTOP) of the cluster and 1500 below. Where there are fewer than 1500 stars above the MSTOP, we take all the stars above the MSTOP and match this number with stars below the MSTOP. This allows us to ensure that we have a reasonable sample of stars on the sub-giant and red-giant branches of the CMD, where most of the information on multiple populations resides.

%%%%%%%%%%%%%%%%%%%%%%%%%%%%%%%%%%%%%%%%%%%%%%%%%%%%%%%%%%%%%%%%%%%%%%%%%%%%%% 
%                                                                                                                                                SINGLE POP
%%%%%%%%%%%%%%%%%%%%%%%%%%%%%%%%%%%%%%%%%%%%%%%%%%%%%%%%%%%%%%%%%%%%%%%%%%%%%%

\section{Results: Single Population Analyses}\label{SinglePop}

For our sample of clusters (NGC 5024, NGC 5272, and NGC 6352), we use the two filter photometry (F606W and F814W) from the ACS Treasury photometry (\citealt{Sarajedini:2007}) with the single population version of BASE-9. We run the single population Bayesian analysis at different helium values (0.23, 0.24, 0.25, and 0.26) in order to study how the results may depend on helium abundance. In this section, we present the prior means and standard deviations, results, and a small discussion for each cluster. We note that for all of the clusters, we removed the horizontal branch stars.

The results of the single population BASE-9 fit to each cluster are given in Tables \ref{NGC5024results} through \ref{NGC6352results} for assumed helium values of 0.23 to 0.26 in 0.01 increments, for both the solar enrichment and enhanced enrichment models. The estimates of each parameter are given by their posterior medians, and the intervals are 90\% Bayesian credible intervals constructed with the 5\% and 95\% posterior quantiles. The MCMC sampling history for NGC 5024 is shown in Figure \ref{5024resultsingle} for the [$\alpha$/Fe] = 0.0 isochrone grid as an example. 

\begin{figure*} 
\epsscale{1.2}
\plotone{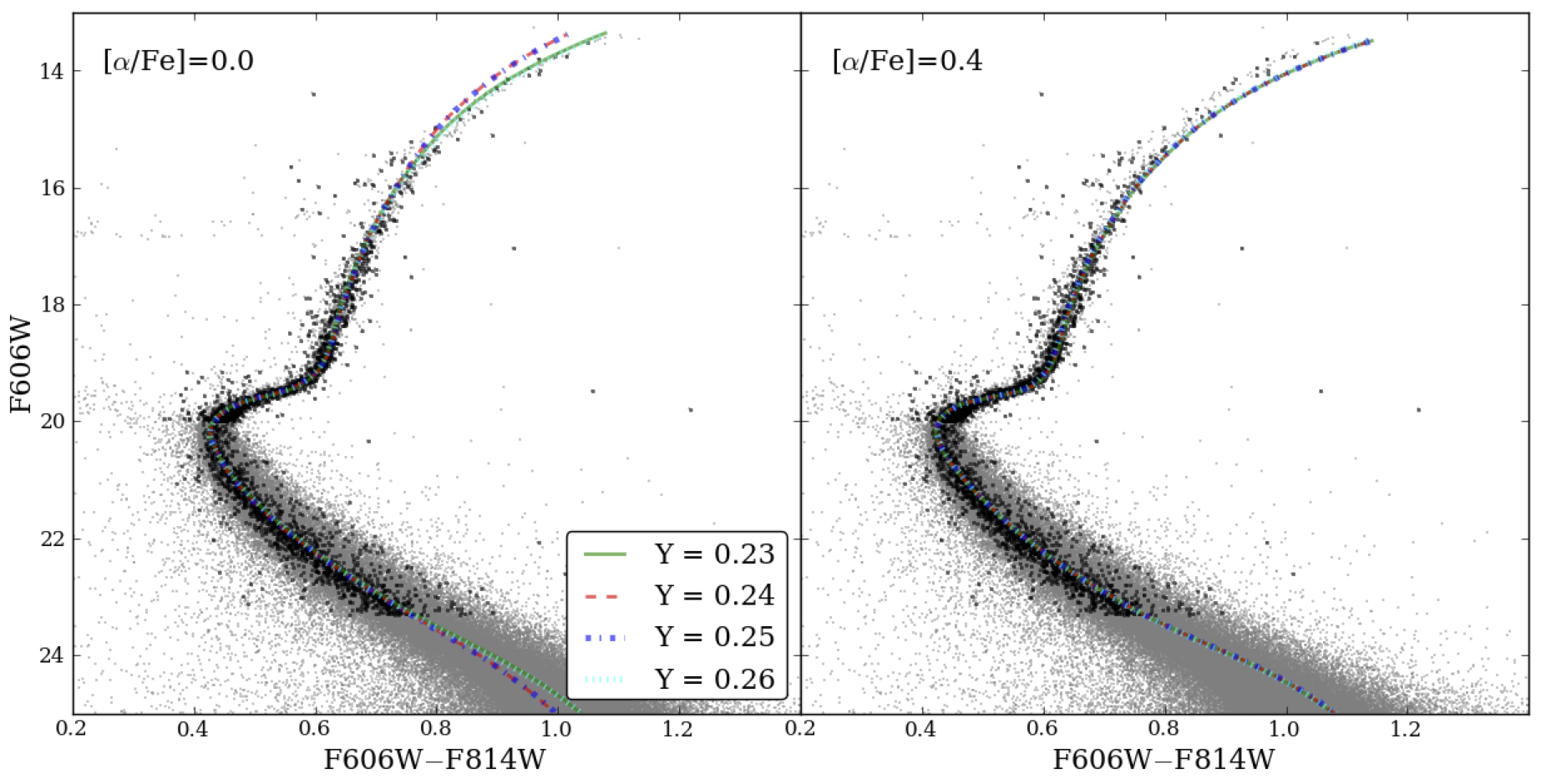}
\caption{F606W and F814W color-magnitude diagrams of NGC 5024 from the ACS treasury data. In each panel, all stars are shown in gray and randomly selected stars used in the fit shown as black dots. The resulting median isochrones determined from the Bayesian analysis are plotted over the stars for the different helium values: solid green line for Y=0.23, dashed red line for Y=0.24, dot-dash blue line for Y=0.25, and a dotted cyan line for Y=0.26. The various fits are often visually indistinguishable. The left panel shows the results with the solar abundance [$\alpha$/Fe]=0 models and the right panel shows the results with the enhanced [$\alpha$/Fe]=0.4 models. For the [$\alpha$/Fe]=0 results in the left panel, the bi-modality of the results can be seen, with the two more metal-rich results (where Y=0.23 and Y=0.26) providing a better visual fit to the RGB.}
\label{5024CMDsingle}
\end{figure*}

\begin{figure*} 
%\epsscale{1.}
\plotone{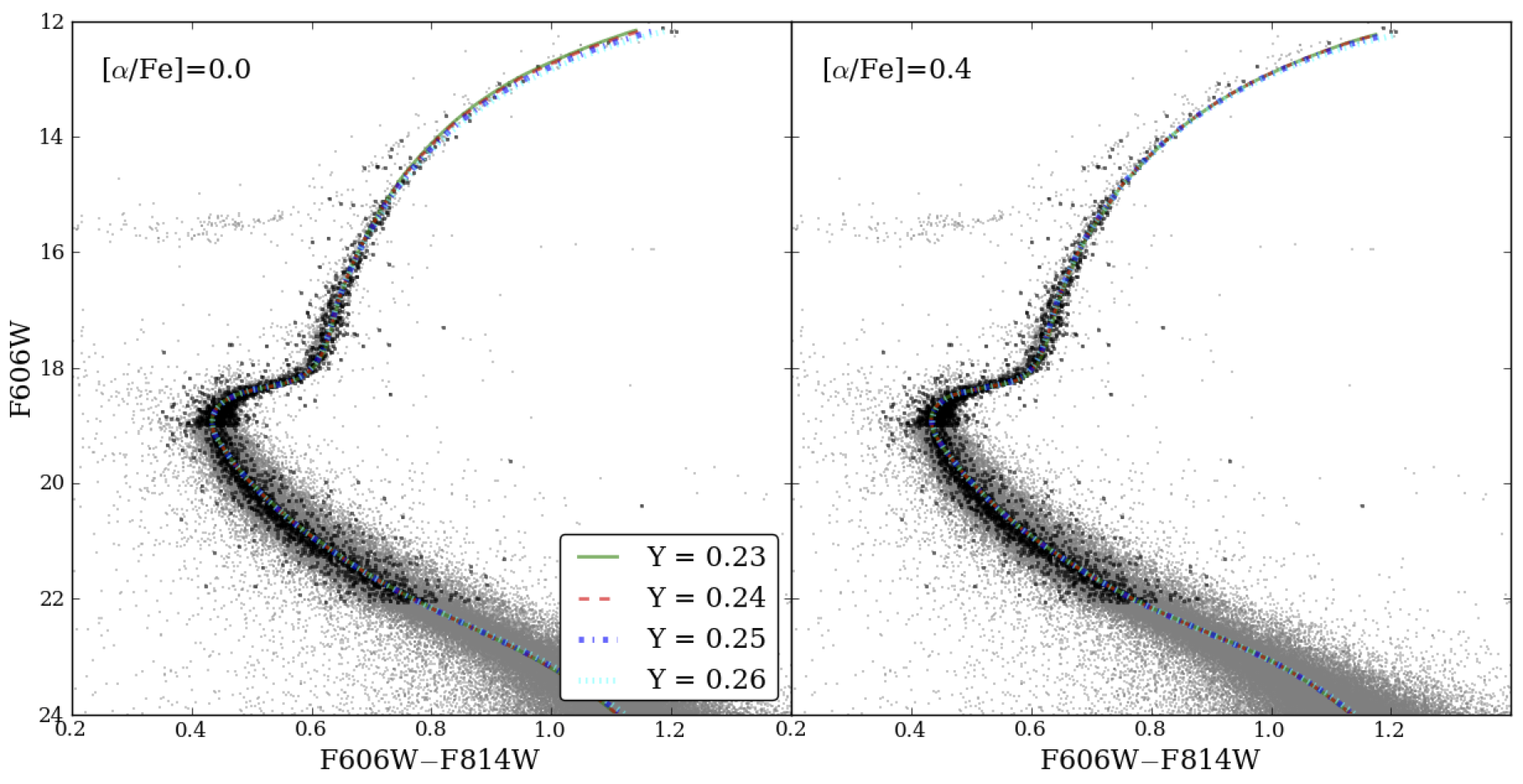}
\caption{Same as Figure \ref{5024CMDsingle}, but for NGC 5272.}
\label{5272CMDsingle}
\end{figure*}

\begin{figure*} 
%\epsscale{1.}
\plotone{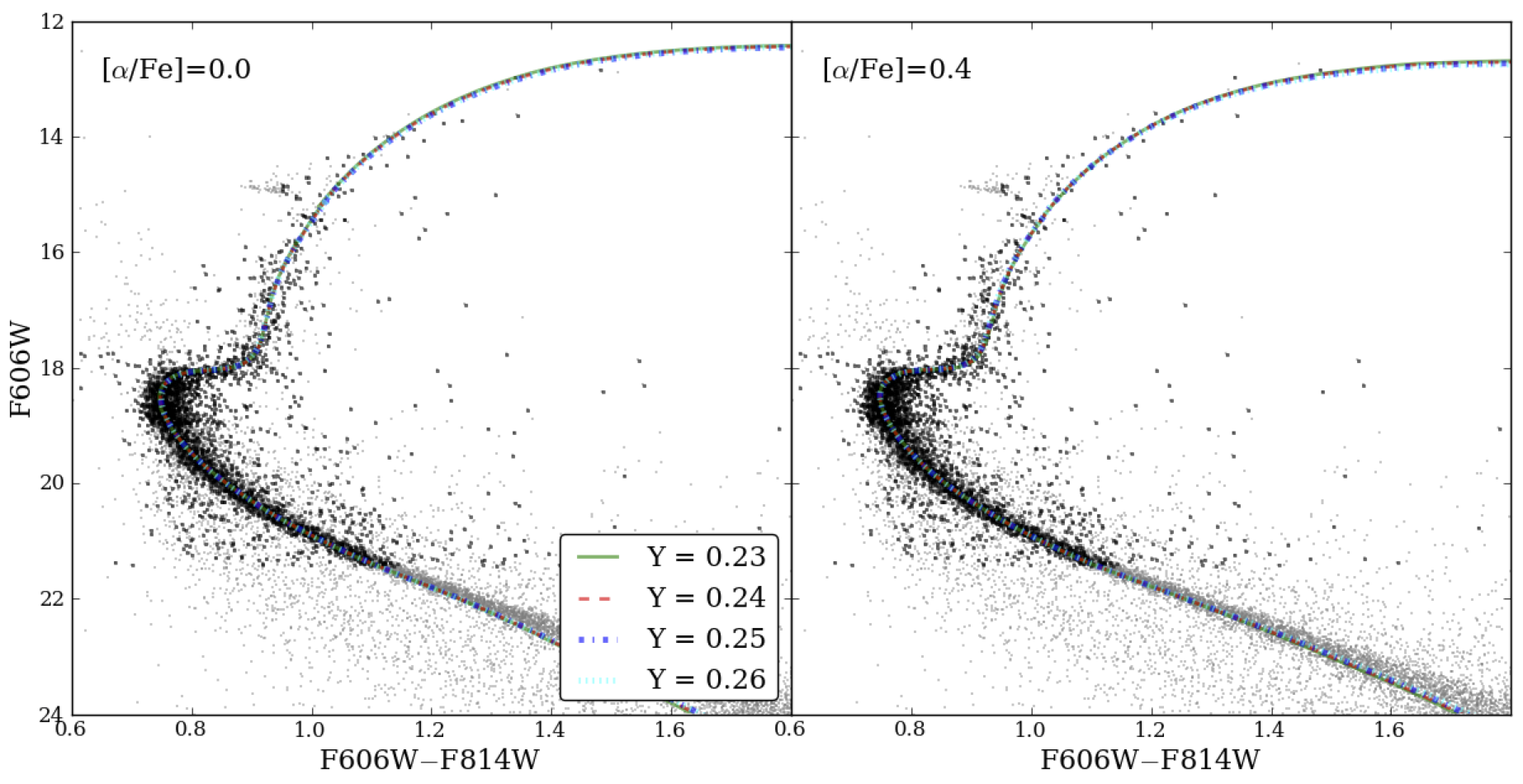}
\caption{Same as Figure \ref{5024CMDsingle}, but for NGC 6352.}
\label{6352CMDsingle}
\end{figure*}

Figures \ref{5024CMDsingle} through \ref{6352CMDsingle} show the cleaned photometry of NGC 5024, NGC 5272, and NGC 6352 as gray points. The sample used in the analysis for each cluster are indicated by the black points, randomly selected as described in Section \ref{Methods}. We use the resulting posterior medians for each parameter to generate the best fit isochrone and plot this on the CMD, as seen in Figures \ref{5024CMDsingle} through \ref{6352CMDsingle}. We list priors from published values by \cite{Harris:1996}, \cite{Harris:2010}, \cite{Dotter:2011}, and \cite{Roediger:2014} in the first two rows of Tables \ref{NGC5024results} through \ref{NGC6352results} for reference.

%%%%% Single 5024 %%%%%%
\subsection{NGC 5024}\label{single5024}

\begin{figure*}
%\epsscale{1.}
\plotone{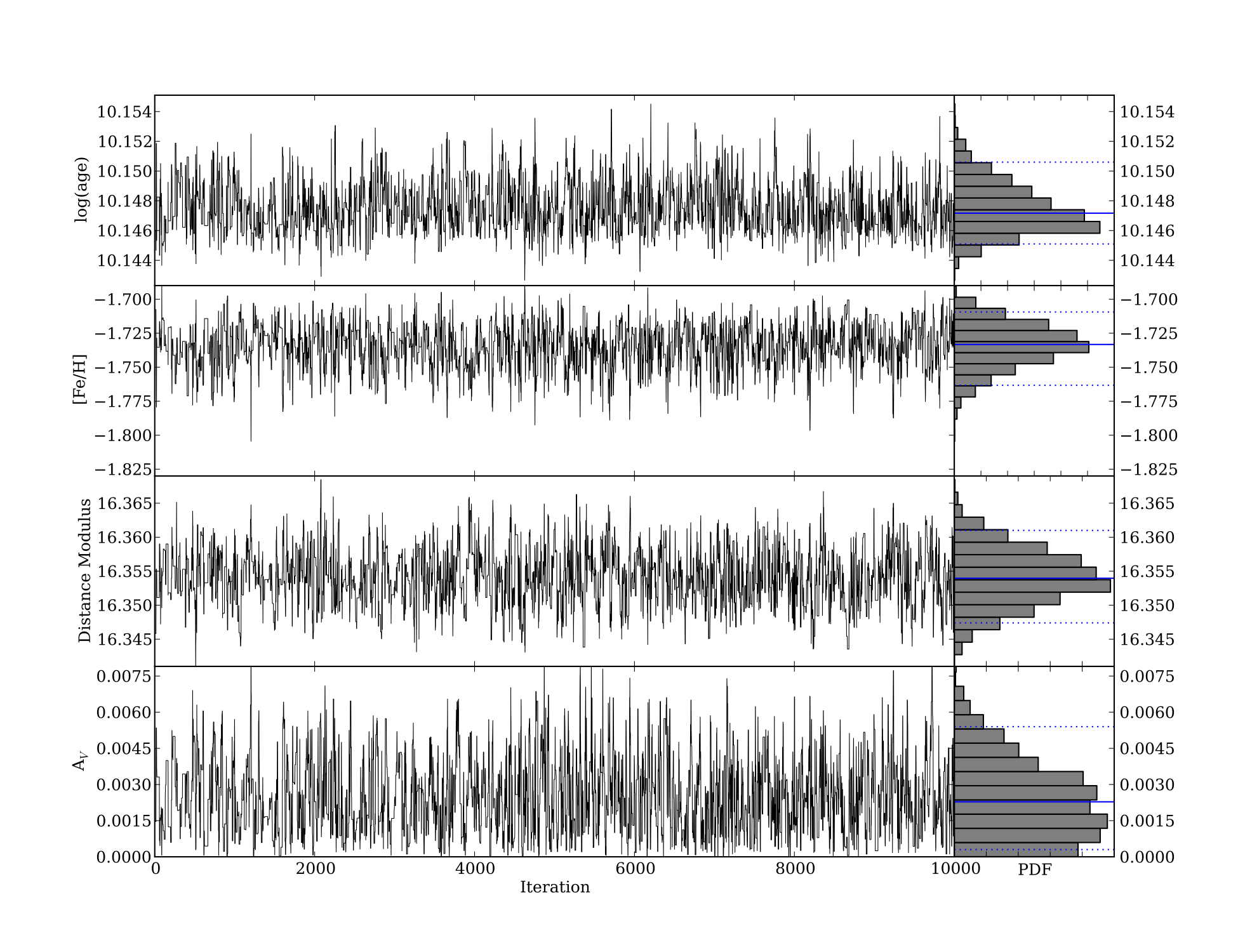}
\caption{Sampling history of BASE-9 fits for NGC 5024 as a single population in the F606W and F814W filters. From top to bottom, the long panels show the sampling chains for log(Age), [Fe/H], distance modulus, and absorption (A$_{V}$). The smaller panels to the right show the probability distribution functions estimated by binning the sampling chains for each parameter. The solid blue line shows the median and the dotted lines show the 90\% Bayesian credible interval.}
\label{5024resultsingle}
\end{figure*}

\renewcommand{\arraystretch}{2}
\begin{table*}
\centering
    \caption{NGC 5024: Single Population}
    \begin{tabular}{@{}ccccc@{}}
    \hline
    \textbf{ } &  \textbf{Age (Gyr)} & \textbf{[Fe/H]} & \textbf{(m--M)$_{V}$} & \textbf{A$_{V}$}  \\
    \hline
Published Value		&	13.25 Gyr\footnotemark[1]	&	--2.10			&	16.32			&	E(B-V)=0.02 \\
Input Prior			&	Uniform 1-15 Gyr	&	--2.10$\pm$0.05	&	16.32$\pm$0.05		&	0.062$\pm$0.02 \\
\hline
[$\alpha$/Fe]=0.0:   & & & &\\
Y = 0.23	&	14.03$^{+0.07}_{-0.11}$	&	-1.733$^{+0.024}_{-0.030}$	&	16.354$^{+0.007}_{-0.007}$	&	0.002$^{+0.003}_{-0.002}$	\\
Y = 0.24	&	14.98$^{+0.07}_{-0.02}$	&	-2.192$^{+0.041}_{-0.035}$	&	16.404$^{+0.009}_{-0.009}$	&	0.033$^{+0.003}_{-0.003}$	\\
Y = 0.25	&	14.89$^{+0.10}_{-0.09}$	&	-2.159$^{+0.037}_{-0.041}$	&	16.394$^{+0.008}_{-0.008}$	&	0.037$^{+0.003}_{-0.003}$	\\
Y = 0.26	&	13.63$^{+0.27}_{-0.16}$	&	-1.696$^{+0.067}_{-0.045}$	&	16.347$^{+0.008}_{-0.007}$	&	0.024$^{+0.003}_{-0.003}$	\\
\hline
[$\alpha$/Fe]=0.4:   & & & & \\
Y = 0.23	&	15.0$^{+0.01}_{-0.01}$		&	-1.668$^{+0.010}_{-0.008}$	&	16.392$^{+0.005}_{-0.006}$	&	0.0$^{+0.001}_{-0.001}$	\\
Y = 0.24	&	15.0$^{+0.01}_{-0.01}$		&	-1.635$^{+0.008}_{-0.009}$	&	16.371$^{+0.005}_{-0.004}$	&	0.0$^{+0.001}_{-0.001}$	\\
Y = 0.25	&	15.0$^{+0.01}_{-0.01}$	&	-1.608$^{+0.012}_{-0.015}$	&	16.354$^{+0.007}_{-0.007}$	&	0.001$^{+0.002}_{-0.001}$	\\
Y = 0.26	&	15.0$^{+0.01}_{-0.01}$	&	-1.614$^{+0.016}_{-0.018}$	&	16.347$^{+0.007}_{-0.007}$	&	0.008$^{+0.003}_{-0.002}$	\\
    \hline
    \end{tabular}
  \footnotetext[1]{\cite{Dotter:2011}}
  \label{NGC5024results}
\end{table*}
\renewcommand{\arraystretch}{1}

Unlike the other two clusters, the results for NGC 5024 using the [$\alpha$/Fe]=0 isochrones appear to be bi-modal for different helium values, with a higher-[Fe/H] mode ($\sim$--1.7) and lower-[Fe/H] mode ($\sim$--2.2), straddling the value we expect based on previous studies ($\sim$[Fe/H]=--2). Visually, the more metal-rich mode, when helium is assumed to be 0.23 or 0.26, appears to be a better fit to the RGB, as seen in Figure \ref{5024resultsingle}. The [$\alpha$/Fe]=0.4 grid results suggest agreement with the more metal-rich mode, although the sampling is restricted due to the edge of the model grid in age.

%%%%% Single 5272 %%%%%%
\subsection{NGC 5272}\label{single5272}

\renewcommand{\arraystretch}{2}
\begin{table*}
\centering
    \caption{NGC 5272: Single Population}
    \begin{tabular}{@{}ccccc@{}}
    \hline
    \textbf{ } &  \textbf{Age (Gyr)} & \textbf{[Fe/H]} & \textbf{(m--M)$_{V}$} & \textbf{A$_{V}$}  \\
    \hline
Published Value		&	12.5 Gyr\footnotemark[1]	&	--1.5				&	15.07			&	E(B-V)=0.01 \\
Input Prior			&	Uniform 1-15 Gyr	&	--1.5$\pm$0.05		&	15.07$\pm$0.05		&	0.031$\pm$0.01 \\
\hline
[$\alpha$/Fe]=0.0:   & & & &\\
Y = 0.23	&	13.37$^{+0.07}_{-0.08}$	&	-1.422$^{+0.008}_{-0.015}$	&	15.064$^{+0.006}_{-0.005}$	&	0.000$^{+0.001}_{-0.000}$	\\
Y = 0.24	&	13.26$^{+0.08}_{-0.08}$	&	-1.385$^{+0.016}_{-0.014}$	&	15.055$^{+0.005}_{-0.005}$	&	0.001$^{+0.002}_{-0.001}$	\\
Y = 0.25	&	13.03$^{+0.10}_{-0.09}$	&	-1.320$^{+0.018}_{-0.017}$	&	15.044$^{+0.005}_{-0.005}$	&	0.000$^{+0.001}_{-0.000}$	\\
Y = 0.26	&	12.79$^{+0.08}_{-0.09}$	&	-1.256$^{+0.014}_{-0.017}$	&	15.034$^{+0.005}_{-0.005}$	&	0.001$^{+0.002}_{-0.000}$	\\
\hline
[$\alpha$/Fe]=0.4:   & & & & \\
Y = 0.23	&	15.00$^{+0.01}_{-0.01}$	&	-1.516$^{+0.008}_{-0.008}$	&	15.092$^{+0.004}_{-0.005}$	&	0.000$^{+0.001}_{-0.000}$	\\
Y = 0.24	&	15.00$^{+0.01}_{-0.01}$	&	-1.495$^{+0.006}_{-0.006}$	&	15.077$^{+0.004}_{-0.004}$	&	0.001$^{+0.001}_{-0.000}$	\\
Y = 0.25	&	15.00$^{+0.01}_{-0.01}$	&	-1.475$^{+0.006}_{-0.007}$	&	15.060$^{+0.005}_{-0.005}$	&	0.001$^{+0.002}_{-0.001}$	\\
Y = 0.26	&	14.99$^{+0.01}_{-0.01}$	&	-1.452$^{+0.005}_{-0.006}$	&	15.041$^{+0.005}_{-0.004}$	&	0.000$^{+0.001}_{-0.000}$	\\
    \hline
    \end{tabular}
  \footnotetext[1]{\cite{Dotter:2011}}
  \label{NGC5272results}
\end{table*}
\renewcommand{\arraystretch}{1}

The results of the Bayesian analysis of NGC 5272 for a single population shows a marked increase of metallicity with an increase in helium fraction for both alpha enrichment scenarios. For this cluster, the spread of assumed helium values leads to a sizable spread in metallicity for [$\alpha$/Fe]=0. When we assume Y = 0.23, we find that [Fe/H] is closest to what we expect based on previous studies. As with NGC 5024, the age is pushed to the edge of the isochrone grid for [$\alpha$/Fe]=0.4.

%%%%% Single 6352 %%%%%%
\subsection{NGC 6352}\label{single6352}

\renewcommand{\arraystretch}{2}
\begin{table*}
\centering
    \caption{NGC 6352: Single Population}
    \begin{tabular}{@{}ccccc@{}}
    \hline
    \textbf{ } &  \textbf{Age (Gyr)} & \textbf{[Fe/H]} & \textbf{(m--M)$_{V}$} & \textbf{A$_{V}$}  \\
    \hline
Published Value		&	13 Gyr\footnotemark[1]	&	--0.70\footnotemark[2]	&	14.43			&	E(B-V)=0.22 \\
Input Prior			&	Uniform 1-15 Gyr	&	--0.70$\pm$0.025		&	14.43$\pm$0.05		&	0.68$\pm$0.23 \\
\hline
[$\alpha$/Fe]=0.0:   & & & &\\
Y = 0.23	&	13.52$^{+0.18}_{-0.32}$	&	-0.589$^{+0.016}_{-0.018}$	&	14.476$^{+0.012}_{-0.016}$	&	0.762$^{+0.006}_{-0.006}$	\\
Y = 0.24	&	13.29$^{+0.19}_{-0.19}$	&	-0.552$^{+0.015}_{-0.015}$	&	14.466$^{+0.011}_{-0.011}$	&	0.761$^{+0.006}_{-0.005}$	\\
Y = 0.25	&	13.02$^{+0.15}_{-0.18}$	&	-0.515$^{+0.012}_{-0.014}$	&	14.456$^{+0.010}_{-0.010}$	&	0.762$^{+0.005}_{-0.005}$	\\
Y = 0.26	&	12.90$^{+0.14}_{-0.14}$	&	-0.500$^{+0.005}_{-0.006}$	&	14.448$^{+0.010}_{-0.010}$	&	0.766$^{+0.005}_{-0.005}$	\\
\hline
[$\alpha$/Fe]=0.4:   & & & & \\
Y = 0.23	&	15.0$^{+0.01}_{-0.01}$	&	-0.852$^{+0.008}_{-0.008}$	&	14.549$^{+0.004}_{-0.004}$	&	0.792$^{+0.003}_{-0.003}$	\\
Y = 0.24	&	15.0$^{+0.01}_{-0.01}$	&	-0.832$^{+0.009}_{-0.008}$	&	14.532$^{+0.004}_{-0.005}$	&	0.791$^{+0.003}_{-0.004}$	\\
Y = 0.25	&	15.0$^{+0.01}_{-0.01}$	&	-0.811$^{+0.012}_{-0.010}$	&	14.515$^{+0.004}_{-0.005}$	&	0.788$^{+0.004}_{-0.004}$	\\
Y = 0.26	&	15.0$^{+0.01}_{-0.01}$	&	-0.789$^{+0.011}_{-0.012}$	&	14.496$^{+0.005}_{-0.004}$	&	0.786$^{+0.005}_{-0.004}$	\\
    \hline
    \end{tabular}
  \footnotetext[1]{\cite{Dotter:2011}}
  \footnotetext[2]{\cite{Roediger:2014}}
  \label{NGC6352results}
\end{table*}
\renewcommand{\arraystretch}{1}

As with NGC 5272, we see a correlation between the assumed helium abundance and [Fe/H]. We also find a sizable difference in the metallicity determined from the two alpha enrichment models. This could plausibly be due to the limit of the isochrones at 15 Gyr, but could point to another route by which assuming values (such as helium) can strongly affect the resulting best fit isochrone. Similar to NGC 5024 for both enrichment models, we find the best agreement in [Fe/H] with the expected value when the helium fraction is assumed to be 0.23.

\subsection{Discussion of Single Population Results}\label{singleDisc}

The errors given in Tables \ref{NGC5024results}, \ref{NGC5272results}, and \ref{NGC6352results} reflect statistical uncertainty, and are not meant to represent astrophysical uncertainty. The uncertainty of the models themselves is unknown, although we do know that the models are unlikely to perfectly reflect true values. Hence, we aim for relative accuracy and precision rather than an absolute accuracy (see \citealt{Stenning:2016}), especially when comparing to previous studies and methods. We also note that the use of different models or different filter combinationsÊcan lead to statistically different results (\citealt{van-Dyk:2009}, \citealt{Hills:2015}).

\begin{figure*} 
\epsscale{1.2}
\plotone{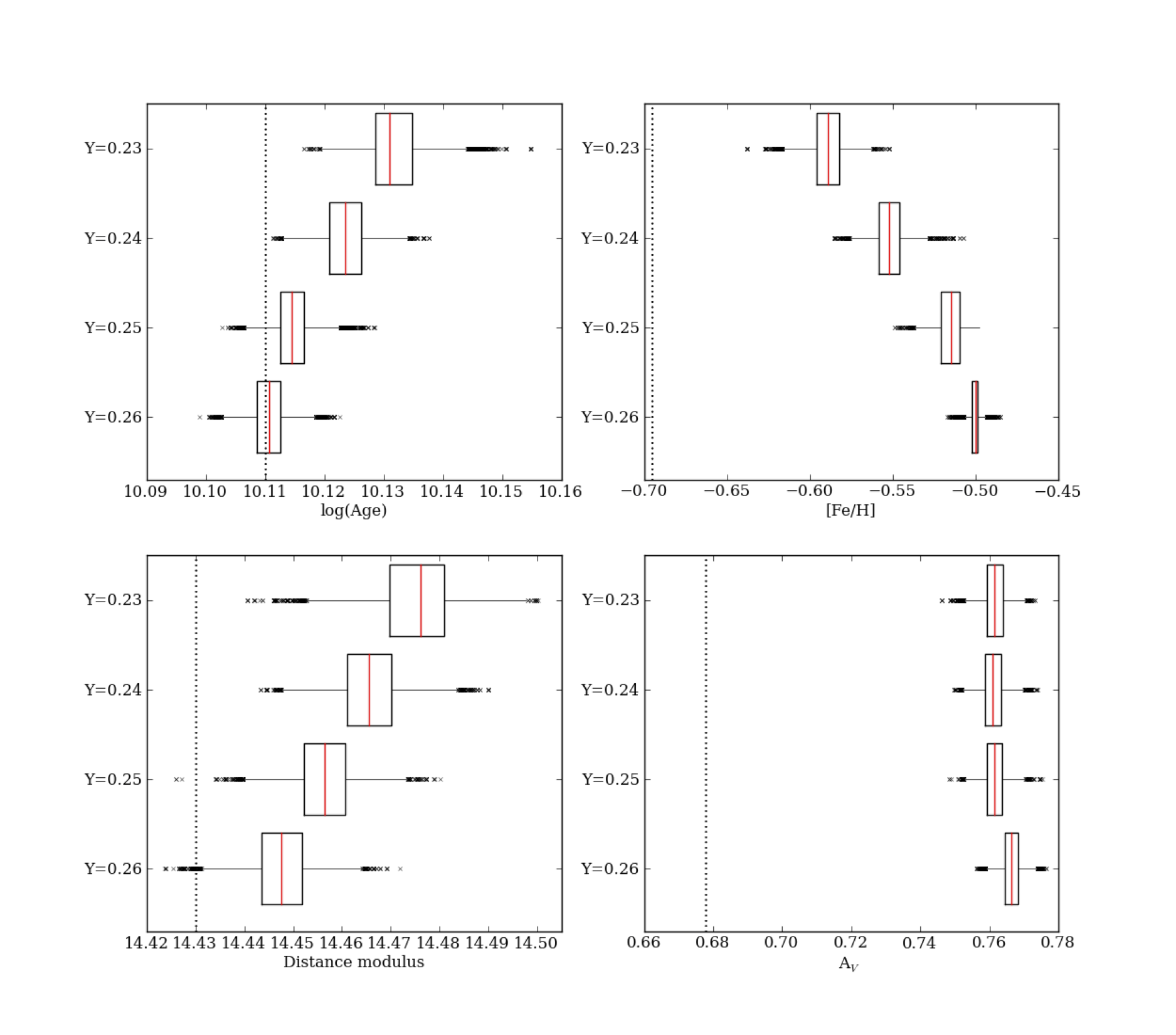}
\caption{A comparison of the single population results from Section \ref{single6352} for NGC 6352 with [$\alpha$/Fe]=0. From top left to bottom right, the panels show the distribution of the sampling history for age, metallicity, distance, and absorption. The y-axis shows the helium value for each box plot in a panel. Published values from Table \ref{NGC6352results} are shown in each panel as the black dotted line.}
\label{boxplot}
\end{figure*}

We find that regardless of the assumed helium abundance, the results tend to reach similar visual fits, but via numerically distinct solutions. Figure \ref{boxplot} provides a comparison of the results from Section \ref{single6352}, the single population runs for NGC 6352 with [$\alpha$/Fe]=0. There is a significant correlation between [Fe/H] and assumed helium abundance, as expected. This trend is seen in the majority of the other single population results as well. We find that even a small change in the helium value ($\Delta$0.03) requires a fairly sizable change in metallicity ($\sim$$\Delta$0.1) in order to maintain the optimal fit to the photometry of NGC 6352. This change can be even more pronounced, as in NGC 5272, where a $\Delta$0.03 change in helium results in an $\sim$$\Delta$0.2 change in [Fe/H].

This implies that the assumptions made about the value of one parameter can strongly affect that of another when only two filters are available. As has been suggested in \cite{Hills:2015}, incorporating additional filters leads to more consistent results. Although at visual wavelengths helium does not strongly affect the morphology of the CMD (making multiple populations difficult to detect), both helium abundance and metallicity affect the RGB shape. Thus, the assumption of a particular ``standard" helium value (which may not reflect the overall helium abundance of the cluster) necessarily affects the resulting metallicity of the adopted best fit isochrone. This suggests that previous studies, if assuming a particular helium value, could bias [Fe/H] values derived from CMD fits towards particular values.

The choice of alpha-enrichment does not strongly affect the distance or absorption results, but affects age significantly, pushing it to the edge of the model grid at 15 Gyr for [$\alpha$/Fe]=0.4. This tends (though not ubiquitously) to push [Fe/H] to more metal-poor values to maintain the best fit isochrone at this older age. In the isochrone morphology, an increase in alpha-enrichment can mimic a decrease in helium abundance. This makes the assumptions of [$\alpha$/Fe] and helium for isochrone fitting doubly important. While we only have two [$\alpha$/Fe] grids available from the Dartmouth isochrones that also include variations in helium, we suggest the effects we observe likely hold for other alpha-enrichment choices as well.

%%%%%%%%%%%%%%%%%%%%%%%%%%%%%%%%%%%%%%%%%%%%%%%%%%%%%%%%%%%%%%%%%%%%%%%%%%%%%% 
%                                                                                                                                                MULTIPLE POP
%%%%%%%%%%%%%%%%%%%%%%%%%%%%%%%%%%%%%%%%%%%%%%%%%%%%%%%%%%%%%%%%%%%%%%%%%%%%%%
 
\section{Results: Two Population Analyses}\label{MultiPop}

We augmented the two-filter photometry for our sample of clusters with new HST UVIS photometry (\citealt{Piotto:2015}). This provides observations spanning five filters: F275W, F336W, F438W, F606W, and F814W. All five filters are used with the two-population BASE-9 analysis. As before, HB stars have been removed as they are not included in the DSED models. In this section, we present the priors, results, and a brief discussion of each cluster.

\begin{figure*}
\epsscale{1.2}
\plotone{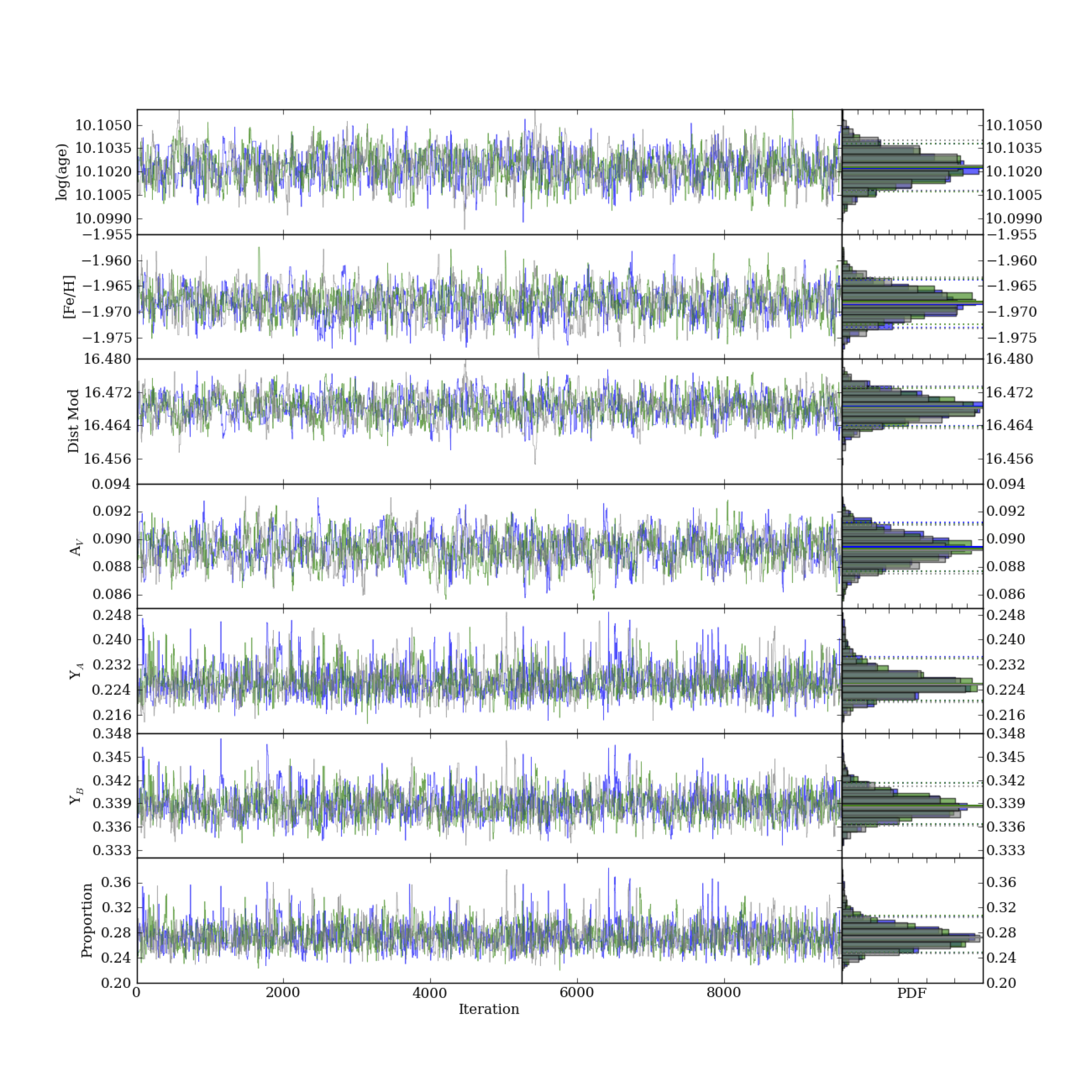}
\caption{The sampling history of the double population BASE-9 fit for NGC 5024 for three chains with random starting values, differentiated by color. For the rows from top to bottom, the long panels show the sampling history for log(Age), metallicity, distance modulus, absorption, helium abundance of population A, helium abundance of population B, and the proportion between the two populations. The histograms to the right of these sampling histories show the posterior distributions of each parameter. The solid line show the medians and the dashed lines indicate the 90\% Bayesian credible intervals for each of the three runs.}
\label{5024resultmulti}
\end{figure*}

The priors we apply are from published values by \cite{Harris:1996}, \cite{Harris:2010}, \cite{Dotter:2011}, and \cite{Roediger:2014}. These are the same priors as discussed above and used in the single-population analysis. For clarity, they are repeated in the first rows of Table \ref{ResultsMulti}. In this table, we also present the results of the two-population BASE-9 fits, given as their posterior medians with the 90\% Bayesian credible intervals. We provide an example of the MCMC sampling history in Figure \ref{5024resultmulti} for three chains of NGC 5024 for the [$\alpha$/Fe]=0 model grid.

\begin{figure*} 
\epsscale{1.25}
\plotone{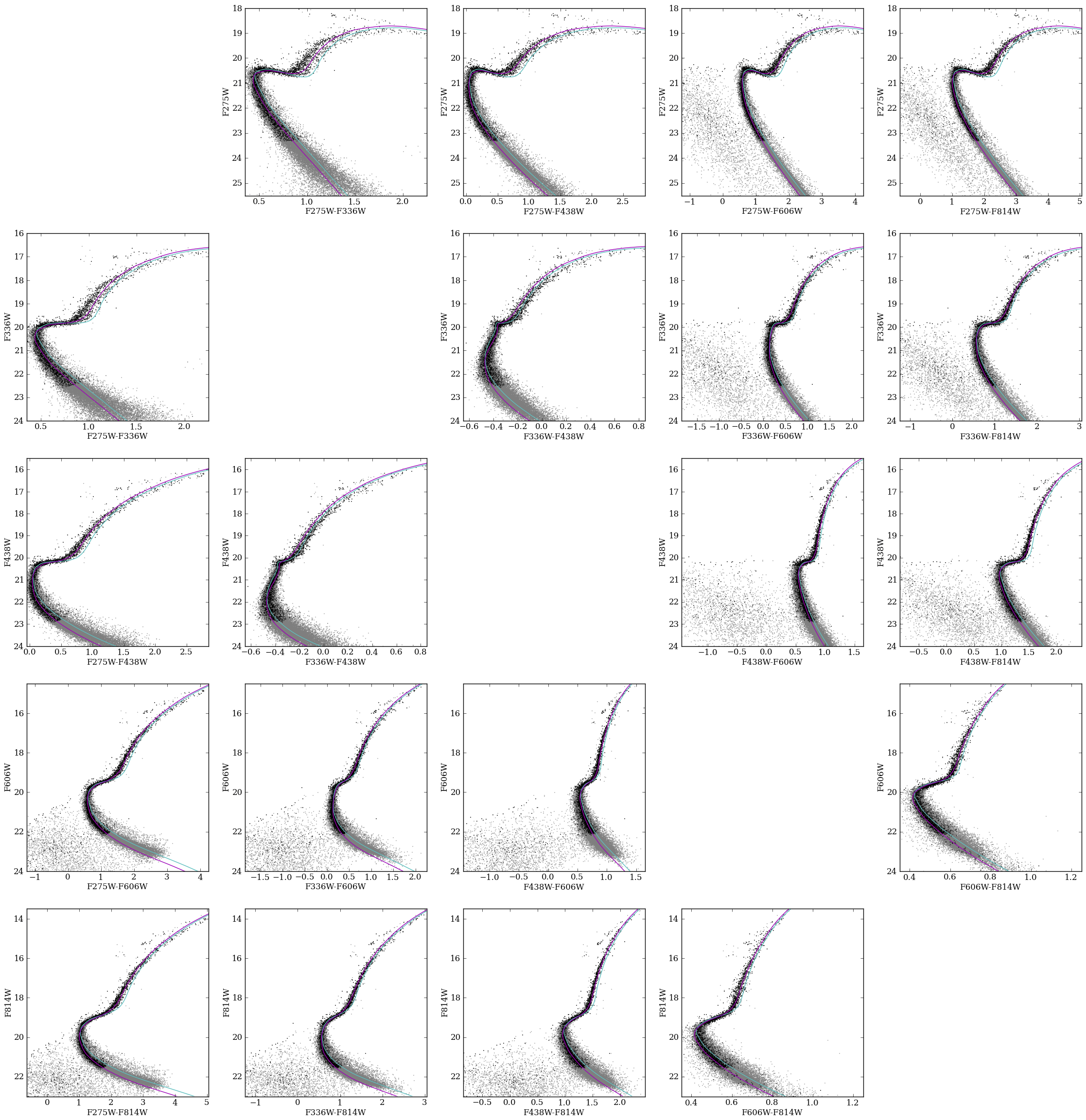}
\caption{A grid of all possible CMDs of NGC 5024 from the five filter UVIS and ACS treasury data (F275W, F336W, F438W, F606W and F814W), wavelength increases moving from left to right and top to bottom. All stars are shown in black and the subsample of stars fit with BASE-9 is shown in gray. The BASE-9 determined model fits are shown as isochrones constructed from median values of the MCMC sampling, with population A shown in cyan and population B in magenta.}
\label{5024CMDmulti}
\end{figure*}

\begin{figure*} 
\epsscale{1.25}
\plotone{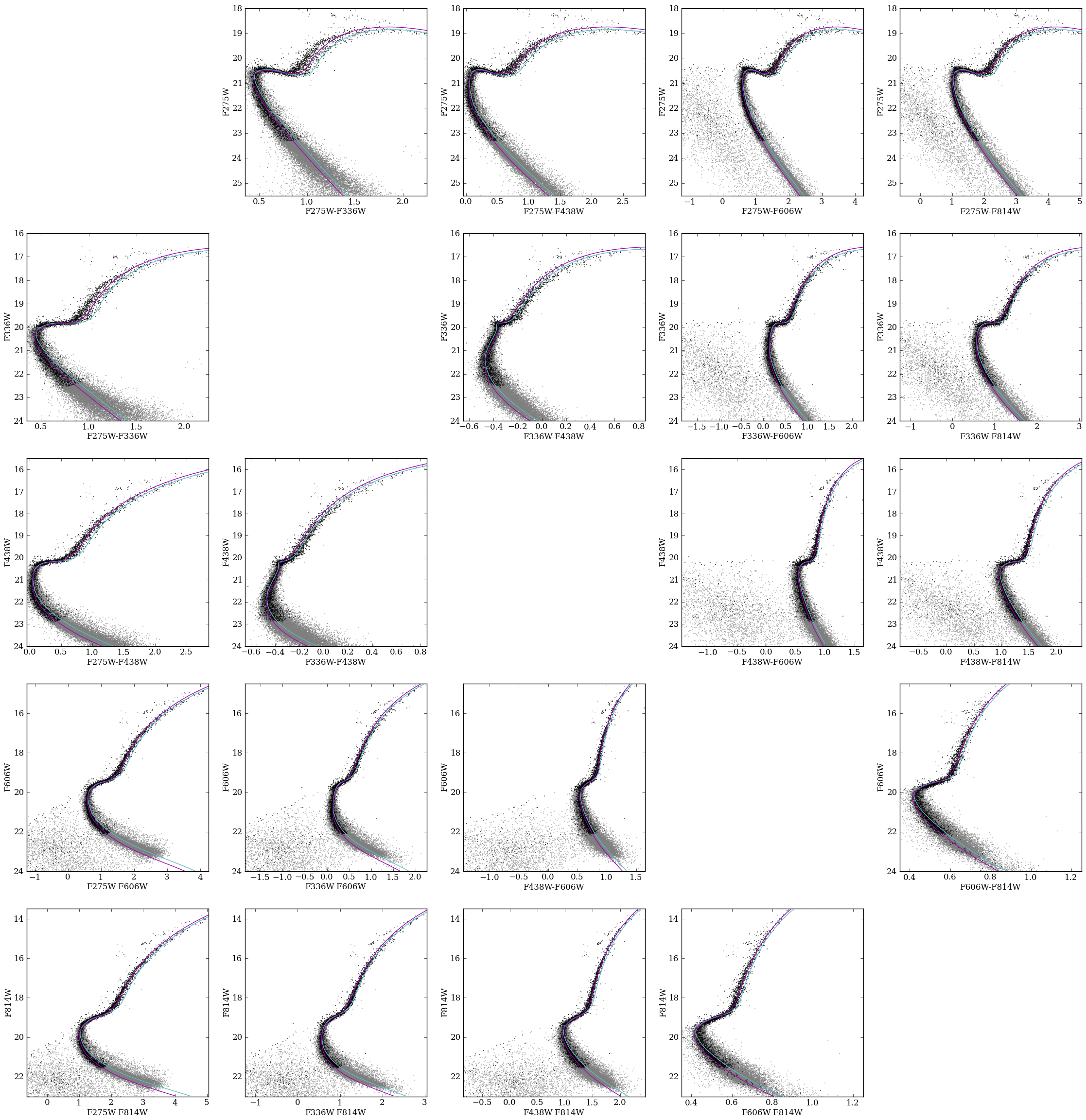}
\caption{Same as Figure \ref{5024CMDmulti}, but for the [$\alpha$/Fe]=0.4 model grid.}
\label{5024CMDmultia4}
\end{figure*}

\begin{figure*} 
\epsscale{1.25}
\plotone{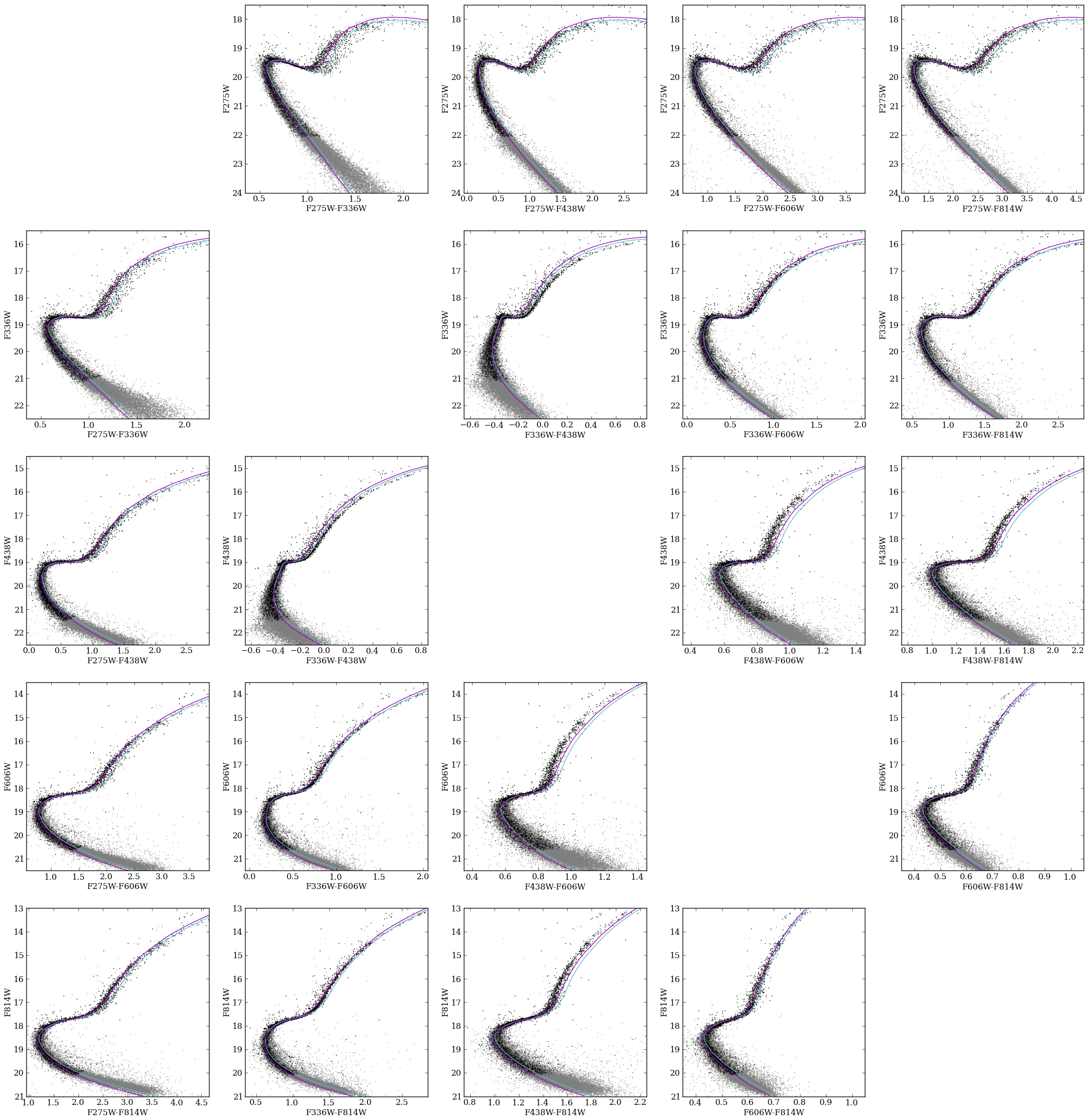}
\caption{Same as Figure \ref{5024CMDmulti}, but for NGC 5272.}
\label{5272CMDmulti}
\end{figure*}

\begin{figure*} 
\epsscale{1.25}
\plotone{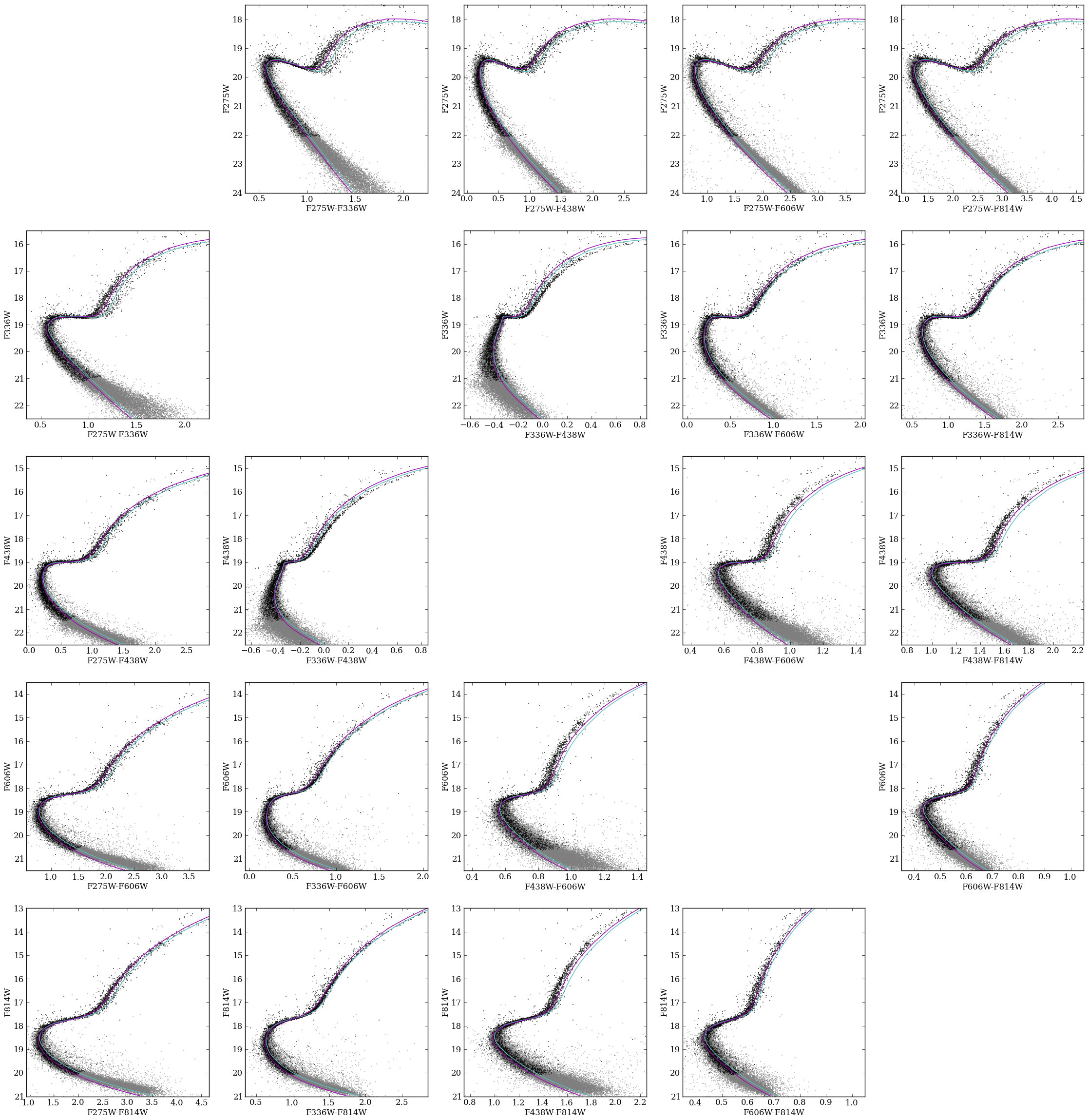}
\caption{Same as Figure \ref{5024CMDmultia4} with the [$\alpha$/Fe]=0.4 model grid, but for NGC 5272.}
\label{5272CMDmultia4}
\end{figure*}

\begin{figure*} 
\epsscale{1.25}
\plotone{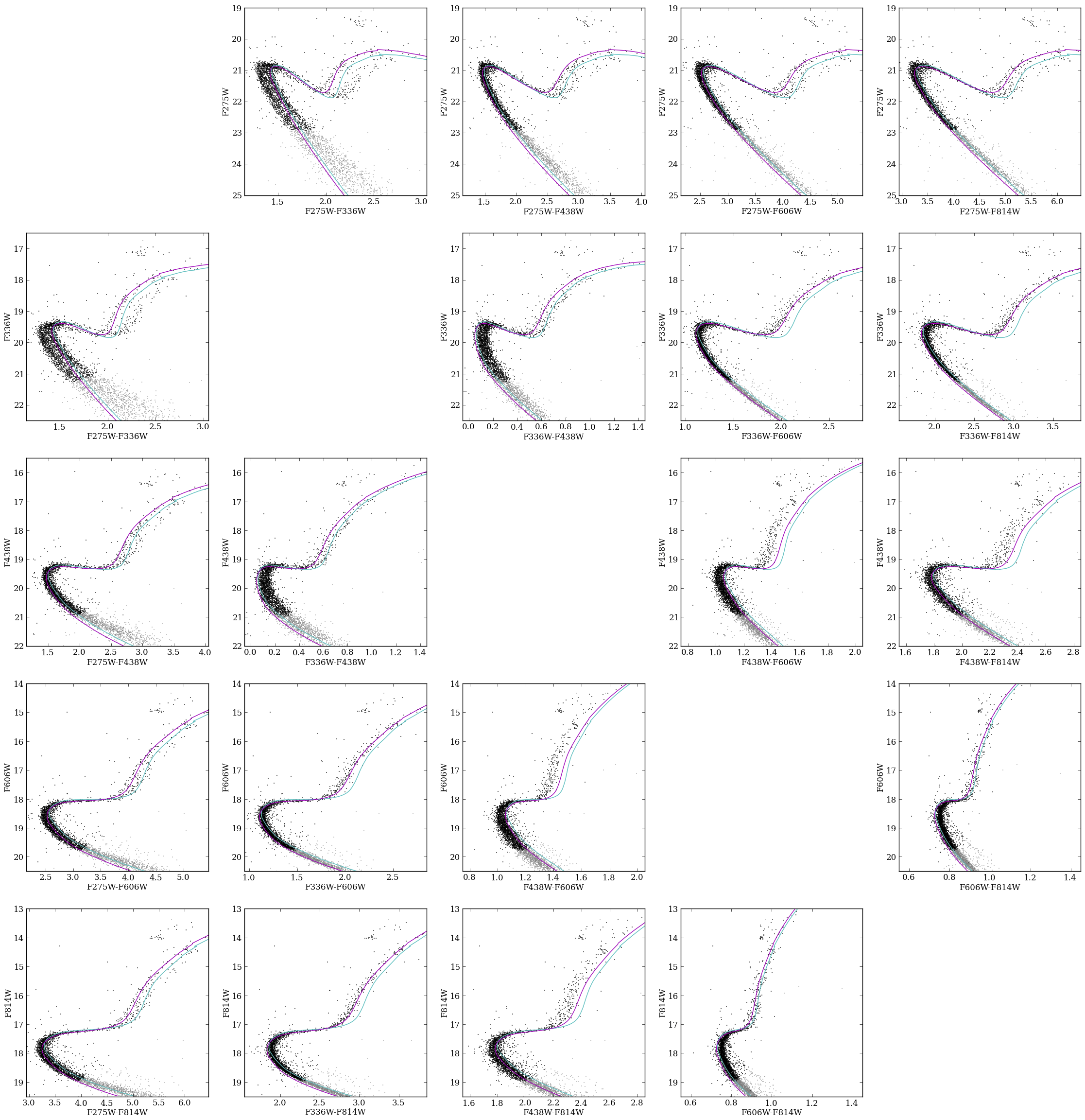}
\caption{Same as Figure \ref{5024CMDmulti}, but for NGC 6352.}
\label{6352CMDmulti}
\end{figure*}

\begin{figure*} 
\epsscale{1.25}
\plotone{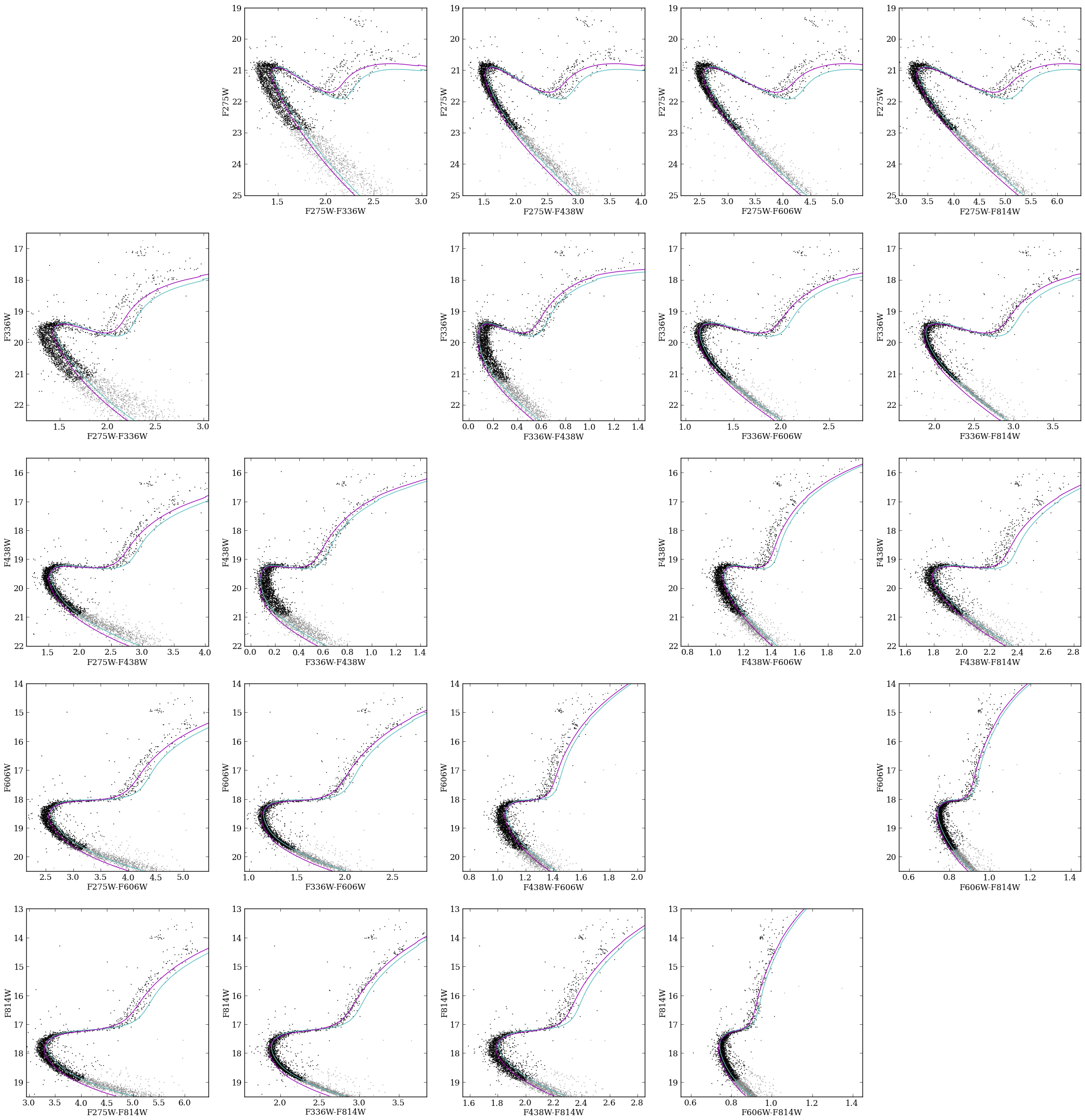}
\caption{Same as Figure \ref{5024CMDmultia4} with the [$\alpha$/Fe]=0.4 model grid, but for NGC 6352.}
\label{6352CMDmultia4}
\end{figure*}

Each panel of Figures \ref{5024CMDmulti} through \ref{6352CMDmultia4} shows the cleaned photometry of NGC 5024 in gray points. As before, the black points represent the sub-sample of stars used in the analysis. We use the resulting posterior medians for each parameter from Table \ref{ResultsMulti} to generate a best fit isochrone for each population in the cluster. These isochrones are plotted on the CMDs for each cluster, as seen in the panels of Figures \ref{5024CMDmulti} through \ref{6352CMDmultia4}.

%%%%%%%%% Two Pop - Observed %%%%%%%%

%%%%% Two pop 5024 %%%%%%
\subsection{NGC 5024}\label{multi5024}

\renewcommand{\arraystretch}{2}
\begin{table*}
\centering
%    \begin{minipage}{180mm}
    \caption{Two Population Results}
    \begin{tabular}{@{}ccccccccc@{}}
    \hline
    
\textbf{ } &  \textbf{Age (Gyr)} & \textbf{[Fe/H]} & \textbf{(m--M)$_{V}$} & \textbf{A$_{V}$} & \textbf{Y$_{A}$} & \textbf{Y$_{B}$} & \textbf{$\Delta$Y} & \textbf{Proportion}  \\
\hline
\textbf{NGC 5024} & & & & & & & & \\
Published Value		&	13.25 Gyr\footnotemark[1]	&	--2.10			&	16.32			&	E(B-V)=0.02 &&&& \\
Input Prior			&	Uniform 1-15 Gyr	&	--2.10$\pm$0.05	&	16.32$\pm$0.05		&	0.062$\pm$0.02 &&&& \\
{[$\alpha$/Fe]}=0.0	& 12.656$^{0.045}_{-0.046}$ & -1.968$^{0.005}_{-0.005}$ & 16.468$^{0.005}_{-0.005}$ & 0.089$^{0.002}_{-0.002}$ & 0.226$^{0.006}_{-0.009}$ & 0.339$^{0.002}_{-0.003}$ & 0.113$^{0.007}_{-0.005}$ & 0.273$^{0.025}_{-0.036}$ \\
{[$\alpha$/Fe]}=0.4	& 14.822$^{0.022}_{-0.023}$ & -2.05$^{0.008}_{-0.008}$ & 16.476$^{0.006}_{-0.005}$ & 0.101$^{0.002}_{-0.002}$ & 0.262$^{0.005}_{-0.005}$ & 0.348$^{0.003}_{-0.003}$ & 0.086$^{0.003}_{-0.003}$ & 0.358$^{0.034}_{-0.036}$ \\
\hline
\textbf{NGC 5272} & & & & & & & & \\
Published Value		&	12.5 Gyr\footnotemark[1]	&	--1.5				&	15.07			&	E(B-V)=0.01 &&&& \\
Input Prior			&	Uniform 1-15 Gyr	&	--1.5$\pm$0.05		&	15.07$\pm$0.05		&	0.031$\pm$0.01  &&&&\\
{[$\alpha$/Fe]}=0.0	& 11.809$^{0.042}_{-0.039}$ & -1.465$^{0.002}_{-0.002}$ & 15.119$^{0.003}_{-0.004}$ & 0.075$^{0.002}_{-0.002}$ & 0.274$^{0.002}_{-0.002}$ & 0.324$^{0.001}_{-0.001}$ & 0.050$^{0.001}_{-0.001}$ & 0.446$^{0.023}_{-0.024}$ \\
{[$\alpha$/Fe]}=0.4	& 14.330$^{0.018}_{-0.02}$ & -1.547$^{0.005}_{-0.004}$ & 15.118$^{0.003}_{-0.004}$ & 0.093$^{0.001}_{-0.001}$ & 0.291$^{0.003}_{-0.002}$ & 0.339$^{0.002}_{-0.002}$ & 0.048$^{0.001}_{-0.001}$ & 0.402$^{0.032}_{-0.025}$ \\
\hline
\textbf{NGC 6352} & & & & & & & & \\
Published Value		&	13 Gyr\footnotemark[1]	&	--0.70\footnotemark[2]	&	14.43			&	E(B-V)=0.22 &&&& \\
Input Prior			&	Uniform 1-15 Gyr	&	--0.70$\pm$0.025		&	14.43$\pm$0.05		&	0.68$\pm$0.23 &&&& \\
{[$\alpha$/Fe]}=0.0	& 11.771$^{0.077}_{-0.081}$ & -0.675$^{0.003}_{-0.004}$ & 14.659$^{0.005}_{-0.004}$ & 0.785$^{0.004}_{-0.004}$ & 0.198$^{0.003}_{-0.003}$ & 0.251$^{0.003}_{-0.003}$ & 0.053$^{0.001}_{-0.001}$ & 0.401$^{0.023}_{-0.026}$ \\
{[$\alpha$/Fe]}=0.4 & 13.971$^{0.094}_{-0.046}$ & -0.728$^{0.011}_{-0.002}$ & 14.618$^{0.002}_{-0.023}$ & 0.825$^{0.007}_{-0.004}$ & 0.279$^{0.009}_{-0.002}$ & 0.328$^{0.004}_{-0.001}$ & 0.049$^{0.002}_{-0.004}$ & 0.355$^{0.013}_{-0.223}$ \\

        \hline
    \end{tabular}
%    \end{minipage}
   \label{ResultsMulti}
   \footnotetext[1]{\cite{Dotter:2011}}
\end{table*}
\renewcommand{\arraystretch}{1}

NGC 5024 has similar metallicity compared to past work, though with a slightly greater absorption and distance. The cluster has a $\Delta$Y of $\sim$0.08 to 0.11 and a proportion around 27\% to 36\%. We find results for absorption, distance, and metallicity that are consistent between the [$\alpha$/Fe] = 0 and [$\alpha$/Fe] = 0.4 models, although the enriched results are a little more metal-poor in [Fe/H], with greater helium.

Our results determine helium values for the second population in NGC 5024 that reach $\sim$0.34. While this is not as high as the 0.42 suggested by \cite{Caloi:2011}, it is still heavily enriched, and our results lend support to their high-He claims. 

While \cite{Caloi:2011} suggest that NGC 5024 is a primarily first population cluster, studies by \cite{DAntona:2008} and \cite{Jang:2014} cite evidence for a heavily second population cluster. Our results indicate that the cluster is relatively balanced in between the two populations, though more second-population dominated.

%%%%% Two pop 5272 %%%%%%
\subsection{NGC 5272}\label{multi5272}

As with NGC 5024, we find that NGC 5272 is slightly younger than suggested in \cite{Dotter:2011}. However, we again recover a similar metallicity, distance, and absorption compared to past work. The $\Delta$Y values are consistent for the two alpha enrichment scenarios, at 0.050 and 0.048. Our analysis suggests NGC 5272 could have $\sim$40 to 45\% first population stars compared to second population stars. This proportion is slightly greater than the 32\% proportion suggested by \cite{Carretta:2009}. Aside from age, we find that our results are consistent between the [$\alpha$/Fe] = 0 and [$\alpha$/Fe] = 0.4 models, though slightly more metal-poor and helium-rich with increased $\alpha$-enhancement.

We determine a helium enhancement of $\sim$0.05 from the first population to the second population of stars in NGC 5272, greater than the suggested $\sim$0 to 0.02 necessary to explain differences in the horizontal branch stars (\citealt{Catelan:2009}, \citealt{Valcarce:2010}, \citealt{Dalessandro:2013}). Our work suggests that an enhancement in helium of 0.02 alone is not sufficient to explain the observed differences of the two populations. However, it may be that our technique is picking up additional information beyond simply helium, as discussed further in Section \ref{Discussion}.

%%%%% Two pop 6352 %%%%%%
\subsection{NGC 6352}\label{multi6352}

We find a younger age than \cite{Dotter:2011}, a metallicity almost equivalent to the spectroscopic value from \cite{Roediger:2014}, and distance and absorptions on par with that of \cite{Harris:1996} and \cite{Harris:2010}. We find a $\Delta$Y of 0.05 to 0.053. Again we see that the higher alpha enrichment requires a slightly lower [Fe/H] and higher helium for both populations.

As with NGC 5272, we find a difference in helium abundances of the populations in the cluster to be slightly greater than that suggested by recent studies (\citealt{Nardiello:2015}). \cite{Nardiello:2015} and this study examine both MS and RGB stars, making for a more direct comparison than with NGC 5272. \cite{Nardiello:2015} assume a metallicity and helium for the first population in order to determine the best helium value for the second population, examining the needed increase in helium in steps to find the best fit to the data. However, our method neither assumes a metallicity nor a helium for the cluster and thus may explain the differences between their results and ours. As we have seen in Section \ref{SinglePop}, assuming one parameter during isochrone fitting can affect the resulting values of other parameters.

%%%%%%%%%%%%%%%%%%%%%%%%%%%%%%%%%%%%%%%%%%%%%%%%%%%%%%%%%%%%%%%%%%%%%%%%%%%%%% 
%                                                                                                                                                DISCUSSION
%%%%%%%%%%%%%%%%%%%%%%%%%%%%%%%%%%%%%%%%%%%%%%%%%%%%%%%%%%%%%%%%%%%%%%%%%%%%%%

\section{Discussion}\label{Discussion}

The results for each cluster, in both the single and two-population fits with [$\alpha$/Fe] = 0.0, are similar to published values. The results from using the [alpha/Fe] = 0.4 models tend to increase helium fractions for both populations and lower [Fe/H]. However, $\Delta$Y remains consistent along with distance, absorption, and (generally) the proportion of the two populations in the clusters. We provide a brief summary of the $\Delta$Y results in Table \ref{DYtable}.

Astrophysically, there is strong motivation to use a two-populations model, particularly when observations at ultraviolet wavelengths are included. Under the two-population Bayesian model, when simulating a single population cluster and analyzing it as a two-population cluster, the proportion either widely varies from 0 to 1 or remains very close to either 0 or 1 (\citealt{Stenning:2016}). We do not see the fitted proportion behave this way for these three clusters, which we take as further evidence that more than a single population is present. Thus, a two population model is more appropriate than a single population for our cluster sample.

Although there are only previous studies of NGC 5272 and NGC 6352 for comparison, it seems that our method tends to require higher $\Delta$Y values to explain differences between populations than other methods. It is possible that the differences between our work and past work arise from the different methodologies used for the analyses, which may focus on stars in different evolutionary stages.

The fits from BASE-9 are partially driven by differences in the red-giant branch, which could be affected by a combination of both changes in helium and in alpha enrichment between the two populations (however, \emph{all} stars in the sample contribute to the fit, regardless of evolutionary stage). We expect oxygen to be depleted in the second population with respect to the first population; the isochrones predict a similar behavior on the RGB for a smaller [$\alpha$/Fe] as for a larger Y. Thus, depleted oxygen in the second population could cause the fit to over-predict Y$_{B}$ and lead to a larger $\Delta$Y. This is motivation to develop our Bayesian analysis to also allow for separate fitting of [$\alpha$/Fe] for each population. %

The accuracy of our analysis depends on NGC 5024, NGC 5272, and NGC 6352 containing two primary populations of stars. As per visual inspection of the CMD, this seems a reasonable assumption for these clusters. However, additional sub-populations may exist that could be detected with a detailed chemical analysis (as in \citealt{Milone:2015} and \citealt{Milone:2015a}). Additionally, we assume that these clusters can be described by a single age and a single metallicity. While we know this is not necessarily the case for all clusters that manifest multiple populations, we believe that any discrepancies present in age and metallicity for these three clusters are likely to be smaller than the current measurement uncertainty of those parameters.

The primary disadvantage of our Bayesian approach is its strict reliance on the accuracy of theoretical isochrones (e.g., \citealt{Dotter:2014}). While ultimately any objective approach must rely on some underlying theory, we feel it is best to do so in an explicit manner and allow the user direct control of relevant inputs (for example, selecting the depth of photometry to use in an isochrone fit). Globular clusters have a rich history of observations in visual filters such as F606W and F814W, and as such the models are reliable and can be fit to the observations with a high level of accuracy. However, in the ultraviolet filters the state of the models lags behind the observations, as high-quality observations at ultraviolet wavelengths (mainly with HST) are just now becoming prevalent. We hope that our results may be able to assist in demonstrating where the models need improvement in order to better represent the observed data. For instance, one common problem we find in our results is the isochrones tend to be too red in certain colors (e.g.: F438W--F606W and F438W--F814W) near the base of the RGB. For NGC 5272 and NGC 6352, this problem persists most of the way up the RGB. From a different perspective, this could be seen as the models predicting a ``longer" SGB (from MSTOP to RGB-TOP) in these colors than is observed in the data.

Another common discrepancy is the lower main sequence, several magnitudes below the turnoff, in the F275W filter (and in some cases, the other two UV filters as well). The observations suggest that the lower main sequence is not as linear as expected, currently limiting the inclusion of the fainter stars in our analysis. For the metal-rich cluster NGC 6352, the RGB shape appears to be have less curvature than predicted by the models, but until other metal-rich clusters are investigated it is unclear if this is a trend. Additionally, combining the ultraviolet filters with the visual filter photometry tends to push the clusters towards younger ages than expected with the [$\alpha$/Fe]=0 model and older than expected with the [$\alpha$/Fe]=0.4.

Hopefully soon, with the current, larger sample of UV observations, updated atmospheric and theoretical models will improve the isochrones and better match the data. Although our methods are limited until such updates are available, we are still able to learn vital information about the two populations in these clusters (particularly helium values, $\Delta$Y, and population proportions) via a rigorous statistical method.

In a future paper, we will apply the two-population Bayesian technique to more clusters. We can then begin to examine the relationship of helium abundances in globular clusters with two populations of stars to other characteristics of the clusters, as well as continuing to provide valuable feedback to the underlying theory. By applying our statistically robust Bayesian analysis method to a larger sample of globular clusters, we will discover more about the characteristics of the globular clusters and gain insight into how these objects formed.

\renewcommand{\arraystretch}{2}
\begin{table}
\centering
%    \begin{minipage}{180mm}
    \caption{Summary of Results}
    \begin{tabular}{@{}cccc@{}}
    \hline
 \textbf{Cluster}	& \textbf{Model} &  \textbf{$\Delta$Y}  &   \textbf{Proportion}  \\
\hline
NGC 5024		&    [$\alpha$/Fe]=0.0		&	0.113$^{0.007}_{-0.005}$		&	0.273$^{0.025}_{-0.036}$    	\\
			&    [$\alpha$/Fe]=0.4		&	0.086$^{0.003}_{-0.003}$		&	0.358$^{0.034}_{-0.036}$   	\\
        \hline
NGC 5272		&    [$\alpha$/Fe]=0.0		&	0.050$^{0.001}_{-0.001}$		 &	0.446$^{0.023}_{-0.024}$  	\\
			&    [$\alpha$/Fe]=0.4		&	0.048$^{0.001}_{-0.001}$		 &	0.402$^{0.032}_{-0.025}$          \\
        \hline
NGC 6352		&    [$\alpha$/Fe]=0.0		&	0.053$^{0.001}_{-0.001}$		&	0.401$^{0.023}_{-0.026}$    	\\
			&    [$\alpha$/Fe]=0.4		&	0.049$^{0.002}_{-0.004}$		&	0.355$^{0.013}_{-0.223}$    	\\
        \hline
    \end{tabular}
%    \end{minipage}
   \label{DYtable}
\end{table}
\renewcommand{\arraystretch}{1}

%%%%%%%%%%%%%%%%%%%%%%%%%%%%%%%%%%%%%%%%%%%%%%%%%%%%%%%%%%%%%%%%%%%%%%%%%%%%%% 
%                                                                                                                                                CONCLUSION
%%%%%%%%%%%%%%%%%%%%%%%%%%%%%%%%%%%%%%%%%%%%%%%%%%%%%%%%%%%%%%%%%%%%%%%%%%%%%%

\section{Conclusion}\label{Conclusion}

The initial results presented in this paper should serve as a proof of concept that we are able to use a Bayesian analysis to identify and characterize two populations in globular clusters. BASE-9 has already been shown to work for single populations (\citealt{von-Hippel:2006}; \citealt{De-Gennaro:2009}; \citealt{Jeffery:2011}; \citealt{Stein:2013}; \citealt{Hills:2015}). Here, we demonstrate that our Bayesian approach can be extended to more than one population, and in particular that it is able to sensitively determine the helium content for two distinct populations.

We find that assuming an [$\alpha$/Fe]=0 enrichment, the clusters NGC 5024, NGC 5272, and NGC 6352 have $\Delta$Y values of 0.048 to 0.113. With an enrichment of [$\alpha$/Fe]=0.4, we observe a range of $\Delta$Y values of 0.050 to 0.079. Additionally, we see the percentage of first population stars in these clusters ranges from approximately 27\% to 45\%.

%%%%%%%%%%%%%%%%%%%%%%%%%%%%%%%%%%%%%%%%%%%%%%%%%%%%%%%%%%%%%%%%%%%%%%%%%%%%%% 
%                                                                                                                                                ACKNOWLEDGEMENTS
%%%%%%%%%%%%%%%%%%%%%%%%%%%%%%%%%%%%%%%%%%%%%%%%%%%%%%%%%%%%%%%%%%%%%%%%%%%%%%
 
\acknowledgments

This material is based upon work supported by the National Aeronautics \& Space Administration under Grant NNX11AF34G issued through the Office of Space
Science, and through the University of Central Florida's NASA Florida Space Grant Consortium. We thank Giampaolo Piotto as PI of the HST UVIS Treasury and Antonino Milone for the data reduction. Work by DS was supported by NSF grant DMS 1208791. DvD was partially supported by a Wolfson Research Merit Award (WM110023) provided by the British Royal Society and by Marie-Curie Career Integration (FP7-PEOPLE-2012-CIG-321865) and Marie-Skodowska-Curie RISE (H2020-MSCA-RISE-2015-691164) Grants both provided by the European Commission.

%%%%%%%%%%%%%%%%%%%%%%%%%%%%%%%%%%%%%%%%%%%%%%%%%%%%%%%%%%%%%%%%%%%%%%%%%%%%%% 
%                                                                                                                                                BIBLIOGRAPHY
%%%%%%%%%%%%%%%%%%%%%%%%%%%%%%%%%%%%%%%%%%%%%%%%%%%%%%%%%%%%%%%%%%%%%%%%%%%%%% 

\newpage
\bibliographystyle{apj}
\bibliography{BASE_2GGC}
\clearpage

%%%%%%%%%% multi pop

% NGC 5024

%alpha=0

%alpha=0.4

% NGC 5272

%alpha=0

%alpha=0.4

% NGC 6352

%alpha=0

%alpha=0.4

\end{document}